\documentclass[prb,aps,superscriptaddress,amsmath,amssymb,floatfix,twocolumn,showpacs,amsfonts,longbibliography] {revtex4-2}
\usepackage{times}
\usepackage[varg]{txfonts}
\usepackage{threeparttable}
\usepackage{textcomp}
\usepackage{graphicx}
\usepackage{subfigure}
\usepackage{subeqnarray}
\usepackage{mathtools}
\usepackage{amsmath}
\usepackage{amssymb}
\usepackage{color}
\usepackage[colorlinks=true,citecolor=blue,urlcolor=blue,linkcolor=blue,hyperindex]{hyperref}
\usepackage{braket}

\begin{document}

\title{Unconventional transport properties in systems with triply degenerate quadratic band crossings}

\author{Zhihai Liu}
\affiliation{College of Physics and Optoelectronic Engineering, Shenzhen University, Shenzhen 518060, China}
\author{Luyang Wang}
\email{wangly@szu.edu.cn}
\affiliation{College of Physics and Optoelectronic Engineering, Shenzhen University, Shenzhen 518060, China}
\author{Dao-Xin Yao}
\email{yaodaox@mail.sysu.edu.cn}
\affiliation{State Key Laboratory of Optoelectronic Materials and Technologies, School of Physics, Sun Yat-sen University, Guangzhou 510275, China}
\affiliation{International Quantum Academy, Shenzhen 518048, China}

\date{\today}

\begin{abstract}
  A quadratic band crossing (QBC) is a crossing of two bands with quadratic dispersion, which has been intensively investigated due to its appearance in Bernal-stacked bilayer graphene. Here, we study an extension of QBCs, the triply degenerate quadratic band crossing (TQBC), which is a three-band crossing node containing two quadratic dispersing bands and a flat band. We focus on two types of TQBCs. The first type contains a symmetry-protected QBC and a free-electron band, the prototype of which is the AA-stacked bilayer square-octagon lattice. In a magnetic field, such a TQBC exhibits an anomalous Landau level structure, leading to a distinctive quantum Hall effect which displays an infinite ladder of Hall plateaus when the chemical potential approaches zero. The other type of TQBC can be viewed as a pseudospin-1 extension of the bilayer-graphene QBC. Under perturbations, this type of TQBCs may split into linear pseudospin-1 Dirac-Weyl fermions. When tunneling through a potential barrier, the transmission probability of the first type decays exponentially with the barrier width for any incident angle, similar to the free-electron case, while the second type hosts an all-angle perfect reflection when the energy of the incident particles is equal to half the barrier height.
\end{abstract}

\maketitle

\section{Introduction}\label{sec:intro}
The band crossings between conduction and valence bands in the electronic band structure of crystalline materials may exhibit some exotic physics, which has stimulated an enormous interest in recent years. The most well-known band crossings are Dirac and Weyl fermions that have been extensively studied at theoretical~\cite{WanXiangang2011RRB,A.A.Burkov2011PRL,Gang.Xu2011PRL,Hongming.Weng2015PRX, S.M.Huang2015NatureC,S.M.Young2012PRL,Zhijun.Wang2012PRB,Zhijun.Wang2013PRB, A.A.Burkov2016NatureM,Armitage2018RMP} and experimental~\cite{Z.K.Liu2014science,B.Q.Lv2015PRX,Su-Yang.Xu2015science, Neupane2014NatureC,Ling.Lu2015science,Yang2015NatureP} levels. One remarkable example is graphene~\cite{Novoselov2004science,Castro2009RMP,Beenakker2008RMP,Goerbig2011RMP}, a single layer of carbon atoms arranged on a honeycomb lattice. In graphene, the unique configuration of carbon lattice generates a linear band crossing at each corner of the hexagonal Brillouin zone (BZ), and the band crossing can be viewed as a massless Dirac fermion with pseudospin-1/2 and described by a $2\times 2$ Dirac Hamiltonian $H=\hbar v_F{\bf k}\cdot\boldsymbol{\sigma}$, where $v_F$ is the Fermi velocity and $\boldsymbol{\sigma}=(\sigma_x, \sigma_y,\sigma_z)$ is a vector of the Pauli matrices~\cite{Palash2011AJP,Semenoff1984PRL}.

Dirac-Weyl (DW) fermions with higher pseudospins may also exist in crystalline materials. For example, pseudospin-1 DW fermions may emerge in some two-dimensional (2D) lattices by fine tunning~\cite{Shen2010PRB,Apaja2010PRA,Bercioux2009PRA,Dora2011RRB, Green2010PRB,WangLuyang2018PRB}. Such a triply degenerate linear band crossing can be described by a $3\times 3$ DW Hamiltonian $H^\prime=\hbar v_F{\bf k}\cdot{\bf S}$, where ${\bf S}=(S_x,S_y,S_z)$ is a vector of the spin-1 matrices satisfying the angular momentum algebra $[S_i, S_j]=i\epsilon_{ijk}S_k$. In comparison with graphene, systems with higher pseudospins may possess some distinctive features. For instance, in graphene, Klein tunneling occurs when the massless Dirac fermions are normally incident to a potential barrier, independent of the barrier width~\cite{Fal'ko2006PRB,Katsnelson2006NatureP,Young2009NatureP,Young2011AnnualR}. In a magnetic field, the unique zero-energy Landau level of graphene leads to an anomalous quantum Hall effect with a half-integer Hall conductivity~\cite{Gusynin2005PRL,ZhangYuanbo2005Nature,K.S.Novoselov2005Nature}. The pseudospin-1 DW fermion, by contrast, has no anomalous quantum Hall effect, since its zero-energy Landau level is non-topological~\cite{Z.Lan2011PRB,Xu.Yong2017PRB}, but presents an all-angle Klein tunneling when the energy of the incident electrons is half the barrier height~\cite{Urban2011PRB}. In addition, systems with pseudospin-1 DW fermions also show particle localization, originating from its flat band~\cite{Apaja2010PRA}.

Besides the linear band crossings, quadratic band crossings (QBCs) have also been investigated~\cite{Sun.Kai2008PRB,SunKai2012NatureP,Y.D.Chong2008PRB,Wang.Zheng2008PRL,FangChen2012PRL}. In fact, in 2D, the QBC is robust if the system hosts time-reversal symmetry and $C_4$ or $C_6$ rotational symmetry~\cite{Sun.Kai2009PRL,Y.D.Chong2008PRB}. A symmetry-protected QBC carries a winding number of 2, and may split into two Dirac points each with winding number 1 when the rotational symmetry is broken down to $C_2$, or three satellite Dirac points each with winding number 1 and a central Dirac point with winding number $-1$ when leaving a threefold rotational symmetry unbroken~\cite{Sun.Kai2009PRL,Mikitik2008PRB,Park2011PRB,deGail2012PRB}. That is, the total winding number is preserved. A typical QBC system is the Bernal-stacked bilayer graphene~\cite{McCann2006PRL,Ohta2006science,Novoselov2006NatureP}, in which a QBC exists at each corner of the hexagonal BZ and can be viewed as a massive chiral fermion. Compared with graphene, the bilayer produces an integer quantum Hall effect owing to the double degeneracy of its zero-energy Landau level~\cite{McCann2006PRL,Novoselov2006NatureP,McCann2006PRB,Castro2007PRL}. In addition, the transmission probability of normally incident electrons in bilayer graphene decays exponentially with the barrier width~\cite{Katsnelson2006NatureP}.
\begin{figure}[t]
  \centering
  \subfigure{\includegraphics[width=1.6in]{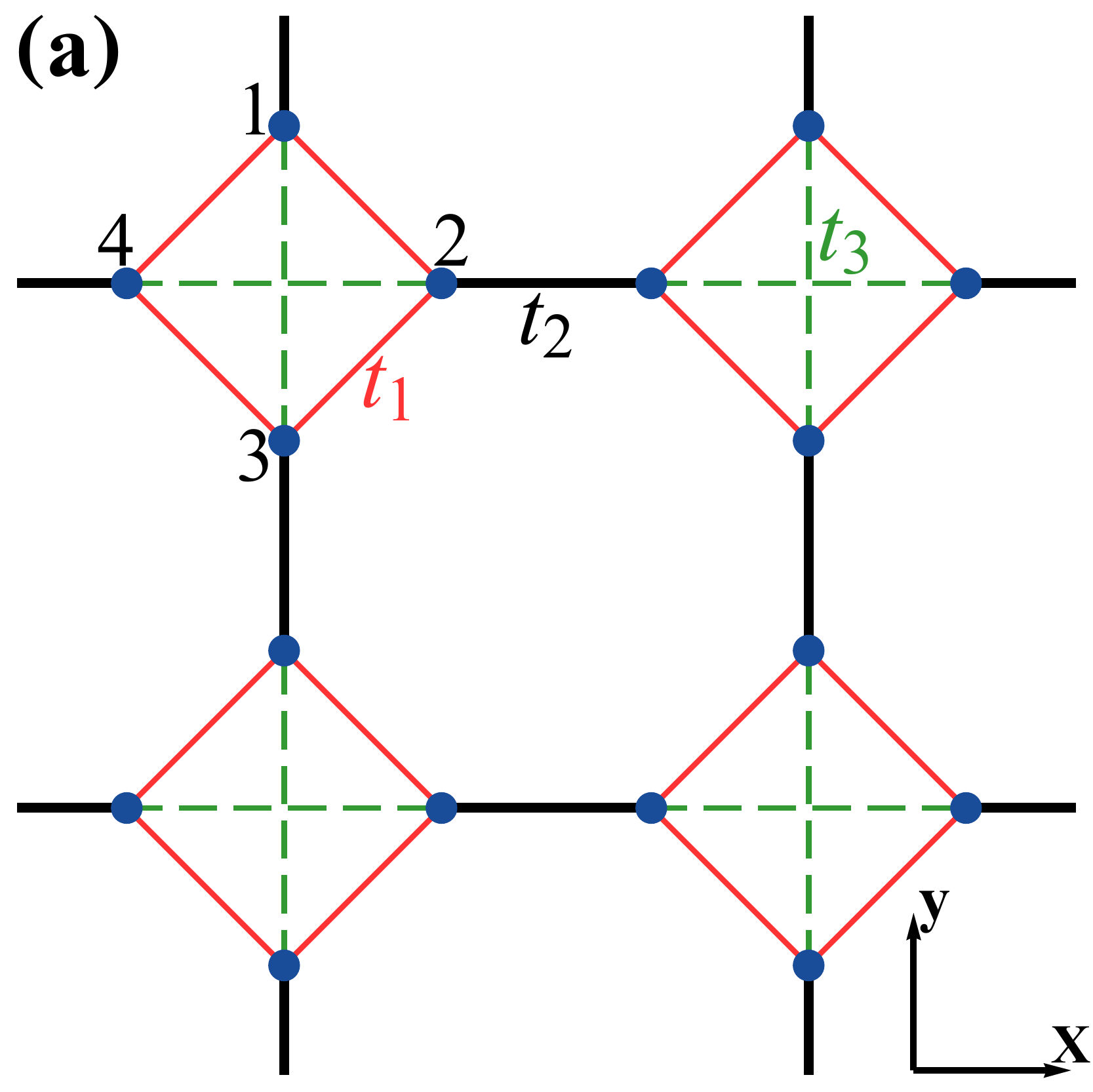}\label{Lattice}}~~
  \subfigure{\includegraphics[width=1.6in]{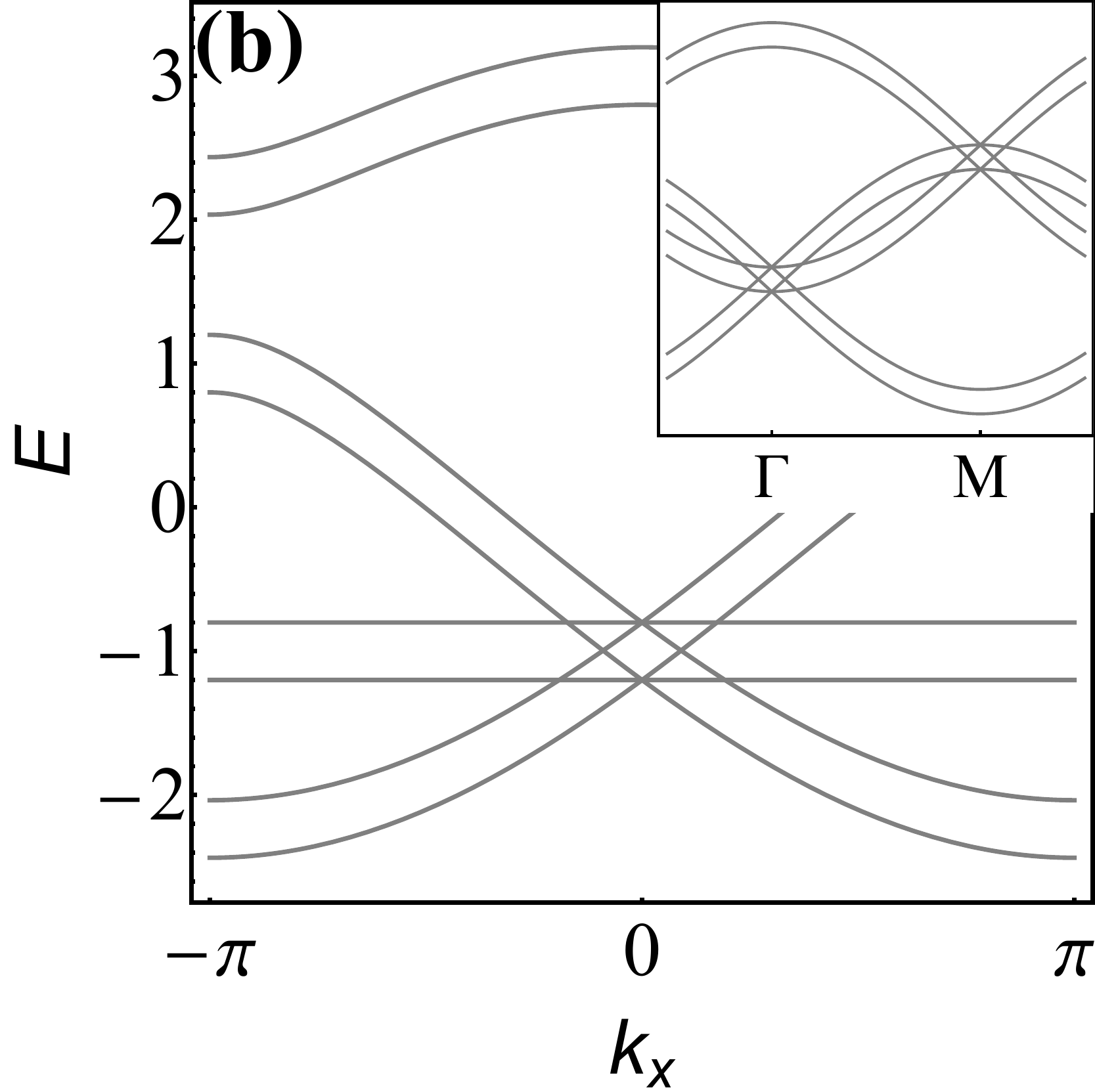}\label{bandb}}
  \subfigure{\includegraphics[width=1.6in]{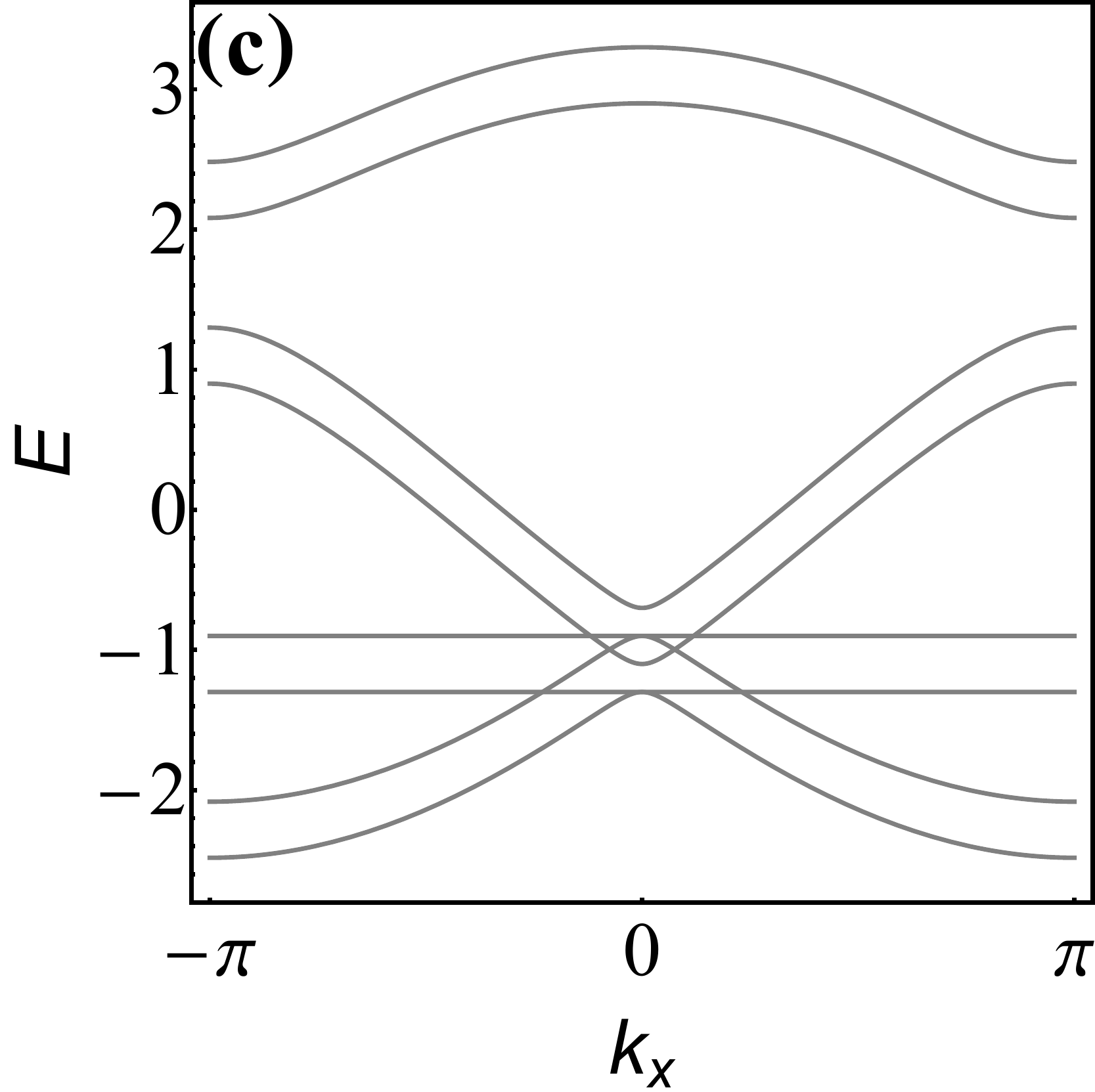}\label{bandc}}~~
  \subfigure{\includegraphics[width=1.6in]{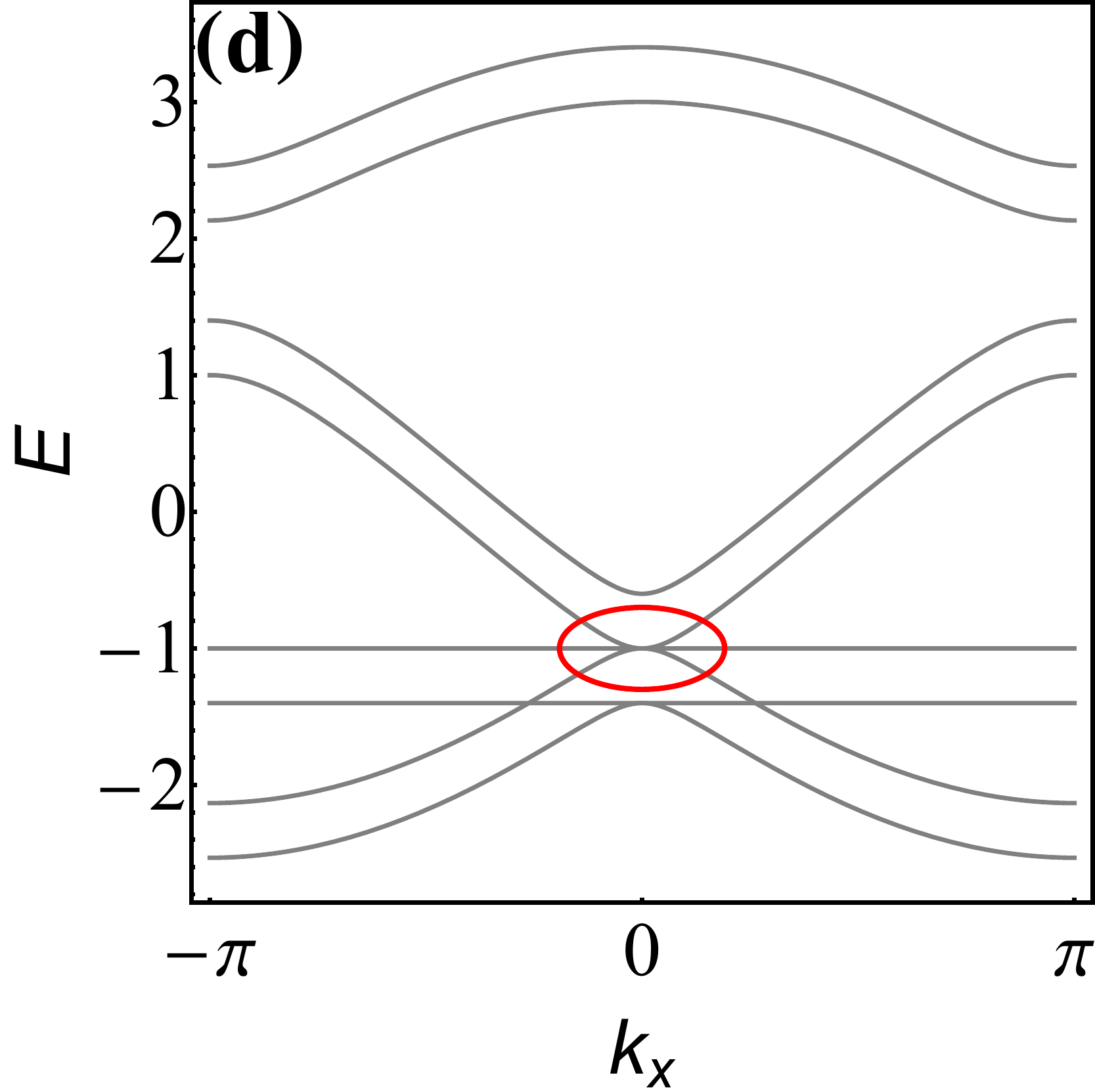}\label{bandd}}
  \caption{(a) The square-octagon lattice. Band structures of the AA-stacked bilayer square-octagon lattice along the $k_x$-axis are shown for (b) $t_2=t_1$ and $t_{\perp}=0.2t_1$, (c) $t_2=1.1t_1$ and $t_{\perp}=0.2t_1$, and (d) $t_2=1.2t_1$ and $t_{\perp}=0.2t_1$. The inset of (b) shows the band structure along $\Gamma$M, and a TQBC can be seen in the red circle in (d). We set $t_1$ as the energy unit.} \label{OctLB}
\end{figure}

Recently, a triply degenerate QBC (TQBC) is discovered in AA-stacked bilayer octagraphene~\cite{Li.Jun2020PRB}, each layer of which is a square-octagon lattice. In this work, we study the band structure of this system, focusing on the TQBC. We find that the TQBC is an accidental touching between a QBC and a free-electron band, which requires fine tuning to occur. We call the band crossing type-I TQBC. The QBC part includes a singular flat band~\cite{Bohm-Jung2021APX}, and is protected by time-reversal and $C_4$ symmetries~\cite{Wei-Feng.Tsai2015NJP}. Different from the case in pseudospin-1 DW fermions~\cite{Chen.Mengsu2012JPCM,Zhihai2021PRB}, the QBC and the free-electron band are decoupled in the TQBC. As a consequence, the barrier tunneling for the TQBC can be broken down into two independent tunneling processes, i.e., tunnelings of QBCs and 2D free electrons. And so is the quantum Hall conductivity. In a magnetic field, the singular flat band of the QBC exhibits an anomalous Landau level structure~\cite{Bohm-Jung2020Nature} and produces an infinite ladder of Hall plateaus in the Hall conductivity. Through a potential barrier, similar to the free-electron scenario, the transmission probability of the TQBC decays exponentially with the barrier width.

We also study another type of TQBCs, which is called type-II TQBCs. The effective model is obtained by extending the effective model for the QBC of the Bernal-stacked bilayer graphene on pseudospins. A type-II TQBC may split into several linear pseudospin-1 DW fermions under perturbations, and presents a similar Hall conductivity as the pseudospin-1 DW fermion system in both the gapless and gapped cases~\cite{Xu.Yong2017PRB}. Remarkably, when tunneling through a sufficiently wide barrier, type-II TQBCs exhibit zero transparency for any incident angle when the energy of incident particles is equal to half the barrier height.

\section{Models}\label{sec:models}
\subsection{Bilayer square-octagon lattice}
The unit cell of the square-octagon lattice contains four sites that form a square as shown in Fig.~\ref{Lattice}. From another point of view, eight sites in the lattice form an octagon, which is analogous to the hexagon in honeycomb lattice. A single layer of carbon atoms arranged on a square-octagon lattice is known as the octagraphene~\cite{Liu.Yu2012PRL,Sheng.Xian-Lei2012JAP}, and high superconducting transition temperatures have been predicted in both the single-layer~\cite{Kang.Yao-Tai2019PRB} and multilayer~\cite{Li.Jun2020PRB} systems. Here, we consider a tight-binding (TB) model in the AA-stacked bilayer square-octagon lattice. The intralayer hoppings $t_1$, $t_2$ and $t_3$ shown in Fig.\ref{Lattice}, and the interlayer nearest-neighbor (NN) hopping $t_\perp$, are considered. Therefore, the TB Hamiltonian reads
\begin{eqnarray}
\mathcal{H}_{TB} = - \sum_{\mathclap{ij,\sigma}}t_{ij} c^{\dagger}_{i\sigma} c_{j\sigma} + \mathrm{h.c.},
\label{Eq.TBHam}
\end{eqnarray}
where $c^{\dagger}_{i\sigma}$ ($c_{i\sigma}$) creates (annihilates) an electron with spin $\sigma$ at site $i$, $t_{ij}=t_1$ for intracell NN hopping, $t_{ij}=t_2$ for intercell NN hopping, $t_{ij}=t_3$ for intracell next-nearest-neighbor (NNN) hopping, and $t_{ij}=t_\perp$ for interlayer NN hopping. The band structures of $\mathcal{H}_{TB}$ can be obtained by diagonalizing the Bloch Hamiltonian in momentum space,
\begin{eqnarray}
H_{bi}({\bf k}) =
\begin{pmatrix}
H_{si}({\bf k}) & t_{\perp}I_{4\times4} \\
t^{*}_{\perp}I_{4\times4} & H_{si}({\bf k})
\end{pmatrix},
\label{Eq.BBHam}
\end{eqnarray}
where $I_{4\times4}$ is the $4\times4$ identity matrix, $H_{si}({\bf k})$ is the Bloch Hamiltonian of a single-layer square-octagon lattice and obtained by taking Fourier transform of TB Hamiltonian (\ref{Eq.TBHam}) with vanishing $t_{\perp}$.

To understand the band structures of the bilayer lattice, we first neglect $t_3$. For a single layer, two pseudospin-1 DW fermions appear, one at the BZ center $\Gamma$(0, 0) and one at the corner M($\pi$, $\pi$) when $t_2=t_1$. Consider a nonzero $t_{\perp}$ for the bilayer lattice. Then two pseudospin-1 DW fermions emerge at $\Gamma$ point and another two at M point when $t_2=t_{1}$, and the two fermions at the same momentum host an energy difference $\Delta E=2t_{\perp}$, as shown in Fig.~\ref{bandb}. When $t_2$ is increased and is away from $t_{1}$ with $t_{\perp}$ fixed, each pseudospin-1 DW fermion is gapped in the way that the triply degenerate point decomposes into a symmetry-protected QBC and a non-degenerate band (see Fig.~\ref{bandc}), with the gap between them increasing with $t_2$. When $t_2$ is increased to $t_1+t_{\perp}$, two TQBCs emerge, one at $\Gamma$ point (shown in Fig.~\ref{bandd}) and the other at M point. Further increasing $t_2$ opens a gap between the original QBC and the non-degenerate band. Therefore, the TQBC is essentially an accidental touching between a QBC and a free-electron band, with the two subsystems uncorrelated. 

With finite $t_3$, for the single-layer lattice, a pseudospin-1 DW fermion is located at $\Gamma$ point or at M point when $t_2=t_1+t_3$ or $t_2=t_1-t_3$, respectively. For the bilayer lattice, there are still two TQBCs located at $\Gamma$ and M when $t_2+t_3=t_1+t_{\perp}$, as was found in Ref.\cite{Li.Jun2020PRB}. The reason is that the pseudospin-1 DW fermion may decompose into a QBC and a non-degenerate band in different ways: a gap can open either between the flat band and the upper band or between the flat band and the lower band, depending on the sign of $t_2-t_{1}$~\cite{Chen.Mengsu2012JPCM}. 

The TQBC in the bilayer square-octagon lattice is named type-I TQBC, and can be described by an effective Hamiltonian
\begin{eqnarray}
H_{\uppercase\expandafter{\romannumeral1}}({\bf k})= - \frac{{\hbar}^2}{4m}
\begin{pmatrix}
k_x^2+k_y^2 & (k_x-ik_y)^2 & 0 \\
(k_x+ik_y)^2 & k_x^2+k_y^2 & 0 \\
0 & 0 & -2(k_x^2+k_y^2)
\end{pmatrix}.
\label{HamA}
\end{eqnarray}
which can be derived by $k\cdot p$ method from the TB model. For simplicity, we have assumed the same effective mass for the upper and lower dispersive bands. Clearly, the model $H_I$ describes two decoupled subsystems, a QBC and a 2D free-electron band. In the QBC, the flat band is singular owing to the discontinuity of its wave function at ${\bf k}=0$. We note that the left upper $2\times2$ Hamiltonian in Eq.~(\ref{HamA}), i.e. $H_{2\times2}\propto d_0I_{2\times2} + d_x \sigma_x+d_y\sigma_y$ with $d_0=k^2_x+k^2_y$, $d_x=k^2_x-k^2_y$ and $d_y=2k_xk_y$, is equivalent to the two-band effective Hamiltonian in Ref.~\onlinecite{Wei-Feng.Tsai2015NJP} via an unitary transformation.

\subsection{Spin-1 extension of bilayer graphene}
The QBC in Bernal-stacked bilayer graphene can be described by an effective Hamiltonian $H_Q({\bf k}) = -\tfrac{{\hbar}^2}{2m_0} \Big[\left(k_x^2-k_y^2\right)\sigma_x + 2k_xk_y \sigma_y \Big]$, where $m_0 \approx 3.04688\times 10^{-13} \mbox{eV/}(\mbox{m/s})^2$ is the effective mass of dispersive quasiparticles~\cite{K.S.Novoselov2005Nature,McCann2006PRL}. Substituting the spin-1 matrices ${\bf S}$ for the Pauli matrices $\boldsymbol{\sigma}$, we get the Hamiltonian for type-II TQBCs,
\begin{eqnarray}
H_{\uppercase\expandafter{\romannumeral2}}({\bf k})=-\frac{{\hbar}^2}{2\sqrt{2}m}
\begin{pmatrix}
m_z & (k_x-ik_y)^2 & 0 \\
(k_x+ik_y)^2 & 0 & (k_x-ik_y)^2 \\
0 & (k_x+ik_y)^2 & -m_z
\end{pmatrix}.
\label{HamB}
\end{eqnarray}
Here we have taken into account a mass term $m_z S_z$, which can gap the TQBC, resulting in Chern numbers $+2$, 0 and $-2$ for the upper, middle and lower band, respectively. In this model, the eigenvector of the flat band is written as
\begin{eqnarray}
{\bf v}_{fb}({\bf k})=\frac{1}{\sqrt{2k^4+m_z^2}}
\begin{pmatrix}
-(k_x-ik_y)^2 \\
m_z \\
(k_x+ik_y)^2
\end{pmatrix}
\label{VFB}
\end{eqnarray}
with $k=\sqrt{k_x^2+k_y^2}$. The flat band is singular for $m_z=0$ and is non-singular for $m_z\neq0$~\cite{Bohm-Jung2019PRB}.

\section{Splitting of the TQBC under perturbations}\label{sec:splitting}
As mentioned in the Introduction, the symmetry-protected QBC may split into two or four Dirac points when the rotational symmetry is broken down to $C_2$ or $C_3$, respectively, but the total winding number is conserved. Take for example the bilayer graphene, the interlayer coupling ${\gamma}_3$ leads to a trigonal warping in its band structure, splitting the QBC three satellite Dirac points each with winding number 1 and a central Dirac point with winding number $-1$. Therefore, the total winding number is $3-1=2$, the same as that of the QBC~\cite{Mikitik2008PRB,deGail2012PRB}. In the effective model of the bilayer-graphene, the splitting of QBC into four Dirac points can be reproduced by considering a perturbation~\cite{McCann2006PRL} $\hbar v_3(k_x \sigma_x-k_y \sigma_y)$, where the velocity $v_3 \propto \gamma_3$, while the splitting into two Dirac points each with winding number 1 can be reproduced by a perturbation $\hbar v_3(k_x \sigma_x + k_y \sigma_y)$.

\begin{figure}[t]
  \centering
  \subfigure{\includegraphics[width=1.6in]{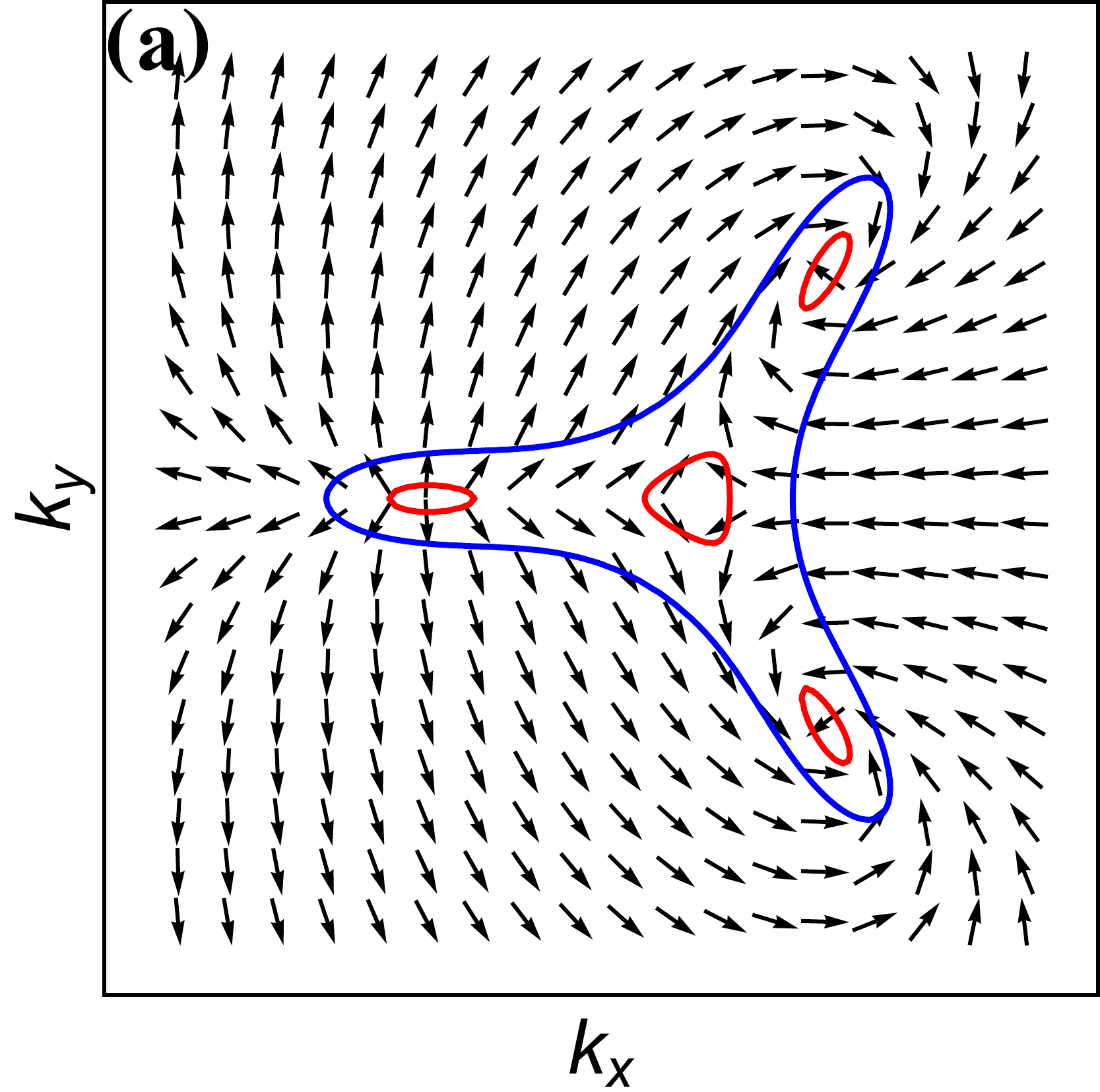}\label{WinF}}~~
  \subfigure{\includegraphics[width=1.6in]{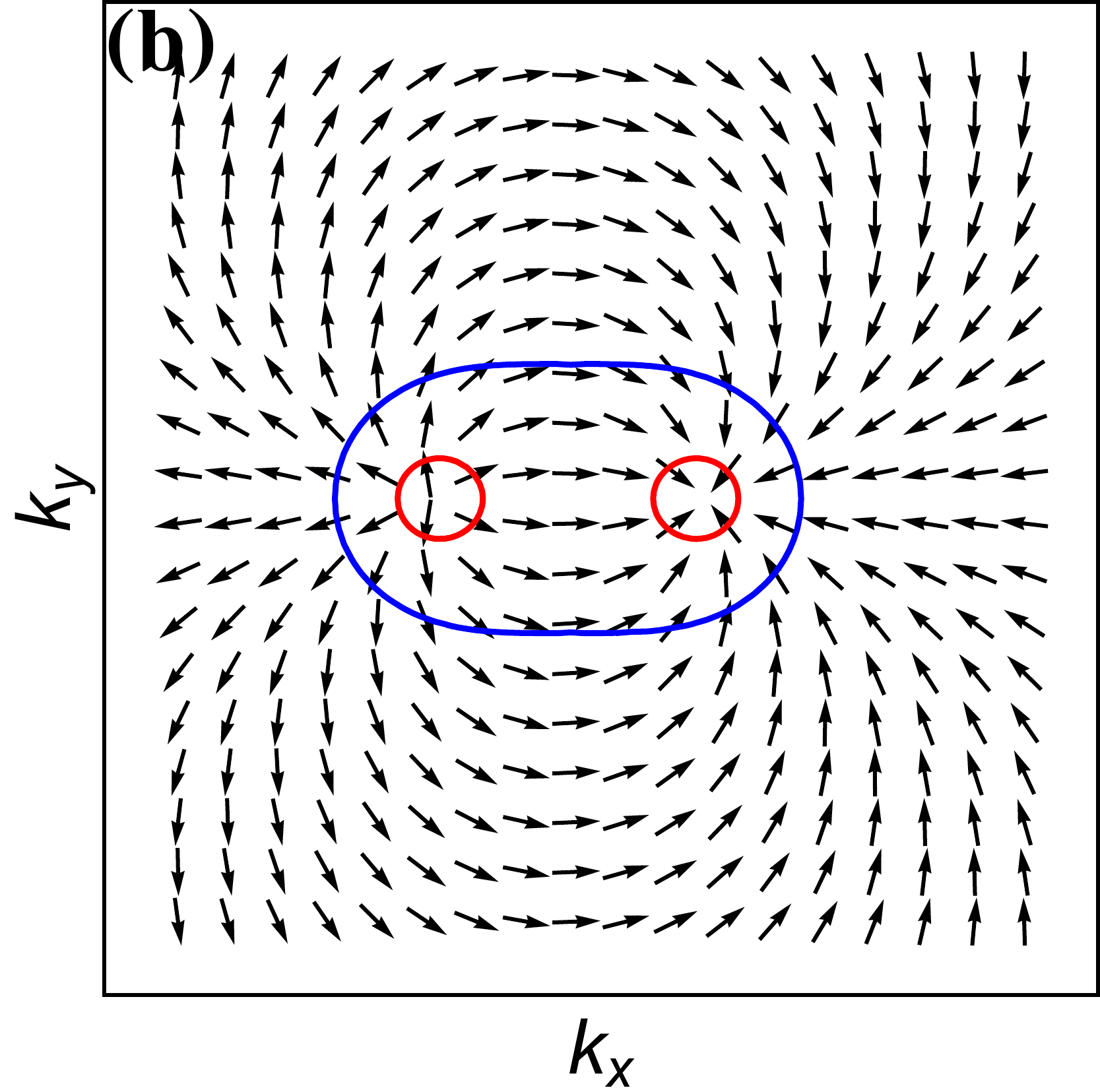}\label{WinT}}
  \caption{The pseudospin textures of the upper band of a type-II TQBC under perturbations (a) $H_F$ and (b) $H_T$. The red lines are equienergy contours at $E=0.075$ and the blue lines at with $E=0.25$. We have set $\hbar=2m=v=1$.} \label{Wind}
\end{figure}

For type-I TQBCs on the square-octagon lattice, when a perturbation breaks the $C_4$ symmetry down to $C_2$, we obtain similar results as for QBCs: two Dirac points with winding number 1 emerge. Additionally, the free-electron band touches the middle band, forming an accidental nodal loop. For type-II TQBCs, a perturbation $H_F=\hbar v(k_x S_x-k_y S_y)$ splits the TQBC into three satellite pseudospin-1 DW fermions located at $(-k_0, 0)$ and $(k_0/2, \pm \sqrt{3}k_0/2)$, and a central pseudospin-1 DW fermion located at $(0, 0)$ (see Fig.~\ref{Wind}), where $k_0=2mv/\hbar$, which resembles the case of bilayer graphene. The type-II TQBC may also split into two pseudospin-1 DW fermions, one located at $(-k_0, 0)$ and the other at $(0, 0)$ under the perturbation $H_T=\hbar v(k_x S_x + k_y S_y)$, also resembling bilayer graphene.

The pseudospin winding number $n_w$ for a node can be defined as the number of rotations ($n_w$ is positive for the counterclockwise rotation) that a pseudospin vector undergoes when the eigenvector rotates one time around the node counterclockwise in the ${\bf k}$-parameter space~\cite{Park2011PRB,Zhihai2021PRB}. $n_w$ is easy to extract from the pseudospin textures of an effective model. As shown in Fig.~\ref{Wind}, the arrows, the mapping vector of pseudospins, are defined as $(\langle u_{{\bf k}\lambda}|S_x|u_{{\bf k}\lambda} \rangle, \langle u_{{\bf k}\lambda}|S_y|u_{{\bf k}\lambda} \rangle)$ where $u_{{\bf k}\lambda}$ is the periodic part of the Bloch wave function of the band $\lambda$. For a type-II TQBC with the perturbation $H_F$, we get the winding number 1 for each satellite node and $-1$ for the central node. On the other hand, under the perturbation $H_T$, both two nodes carry the winding number 1. As we can see in Figs.~\ref{WinF} and~\ref{WinT}, the winding number is 2 along each blue contour. We conclude that the total winding number is conserved for both two perturbations.

It is well-known that the Berry phase has a $2n\pi$ uncertainty since a gauge transformation can be made to the Bloch wave functions,
\begin{equation}
|u_{{\bf k}\lambda} \rangle \to e^{in\zeta({\bf k})}|u_{{\bf k}\lambda} \rangle
\label{GaugeT}
\end{equation}
with $\zeta({\bf k})=\arg(k_x+ik_y)$. In Ref.~\onlinecite{Mikitik2008PRB}, the $2n\pi$ uncertainty is removed by a weak spin-orbit interaction, yielding the Berry phase $\pi$ for each satellite Dirac point and $-\pi$ for the central Dirac point. However, for model $H_{II}$, we directly calculate the Berry phase with respect to the band $\lambda$ via $\gamma_\lambda=\oint_C \mathcal{A}_{\lambda\lambda}({\bf k}) \cdot d{\bf k}$, where $\mathcal{A}_{\lambda\lambda}({\bf k})$ is the Berry connection. Eventually, under gauge transformation~(\ref{GaugeT}), we obtain the Berry phase $2n\pi$ with respect to the upper band of the type-II TQBC. While when considering the perturbation $H_F$, the integral gives the Berry phase $2\pi$ for contours surrounding each of the three satellite pseudospin-1 DW fermions and $2(n-3)\pi$ for a contour around the central node. And under the perturbation $H_T$, the Berry phase is $2\pi$ for the satellite pseudospin-1 DW fermion and $2(n-1)\pi$ for the central fermion. In other words, for both the two perturbations, when taking gauge transformation~(\ref{GaugeT}) for the Bloch wave function of Hamiltonian~(\ref{HamB}), the $2n\pi$ uncertainty of Berry phase arises only on the central node. A similar scenario can also be seen in graphene~\cite{Park2011PRB}.

\section{Landau level structure and Unconventional Quantum Hall effect}\label{sec:LLs}
\subsection{Landau level structure}
We calculate the Landau levels (LLs) for the two types of TQBCs. Taking a vector potential ${\bf A}({\bf r})=(-By, 0, 0)$ that generates a homogeneous magnetic field ${\bf B}$ along $z$-direction. In a magnetic field, the canonical momentum should be replaced as ${\bf P}\to {\bf \Pi}={\bf P}+e{\bf A}({\bf r})$ where ${\bf P}=\hbar{\bf k}$. We introduce ladder operators $\hat{a}=\tfrac{l_B}{\sqrt{2}\hbar}(\Pi_x-i\Pi_y)$ and $\hat{a}^\dagger=\tfrac{l_B}{\sqrt{2}\hbar}(\Pi_x+i\Pi_y)$ with the magnetic length $l_B=\sqrt{\hbar/e|B|}$, which satisfy the commutation relation $[\hat{a},\hat{a}^{\dagger}]=1$.
\begin{figure}[t]
  \centering
  \subfigure{\includegraphics[width=1.6in]{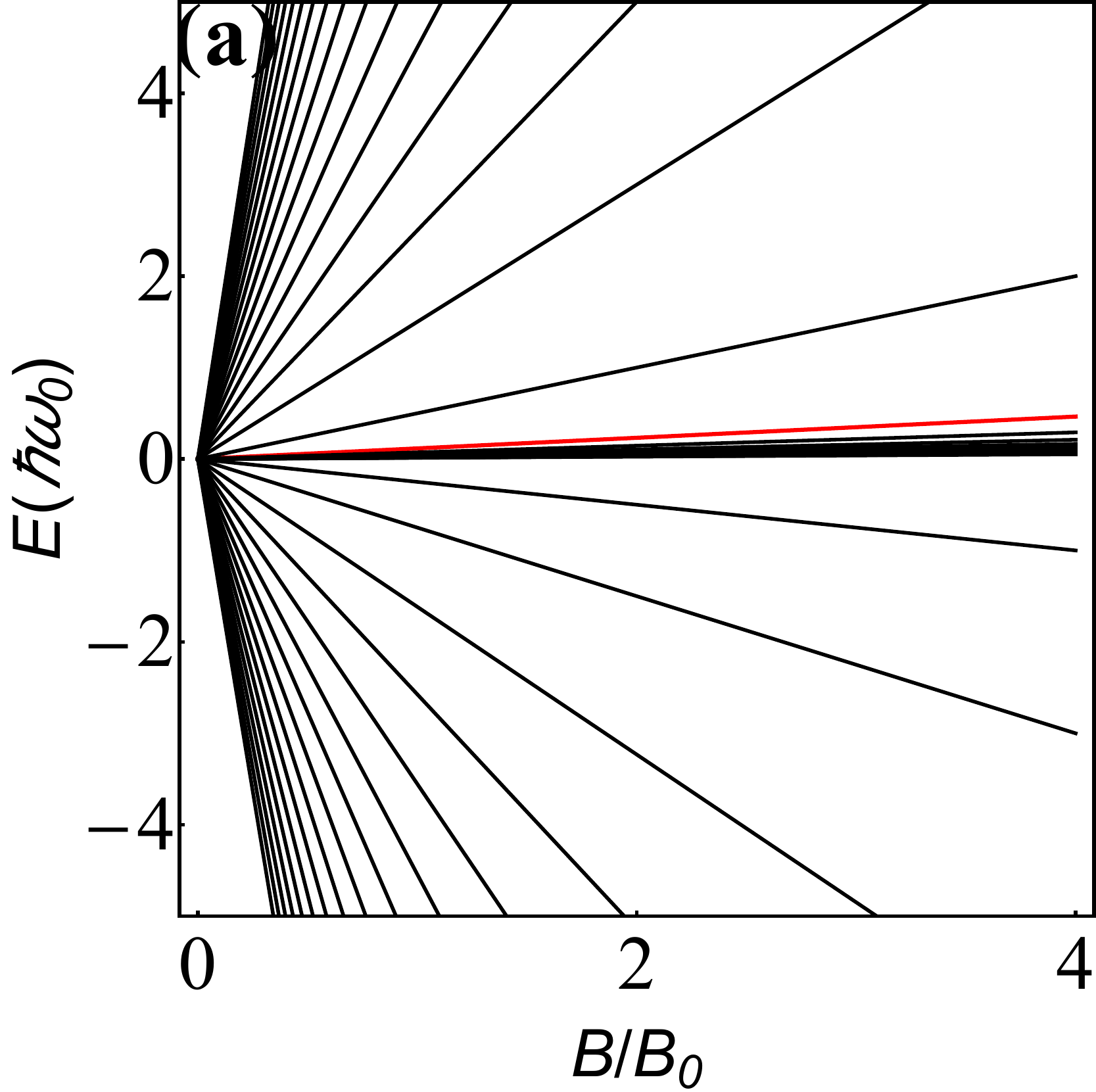}\label{LLsa}}~~
  \subfigure{\includegraphics[width=1.6in]{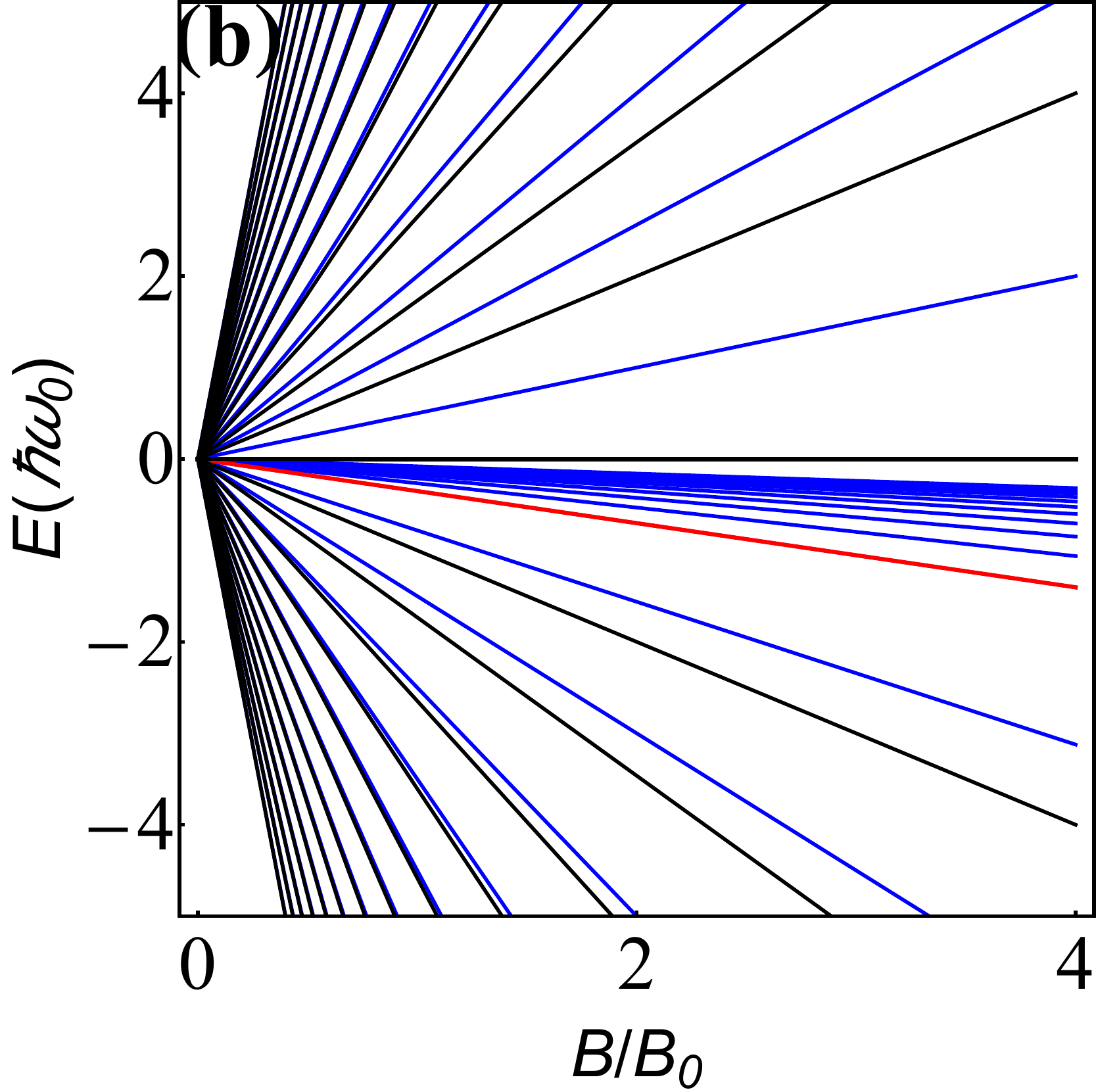}\label{LLsb}}
  \caption{The Landau levels as a function of the magnetic field $B$ for type-I (a) and type-II (b) TQBCs, where $\omega_0=eB_0/m$ with any specified magnetic field $B_0$. The red lines indicate the $n=2$ levels in $\lambda=0$ groups. In (b), the black lines correspond to $\delta=0$ and the blue lines correspond to $\delta=0.5$.} \label{LLs}
\end{figure}

We solve the eigenvalue equation $H^{\bf B}\psi_{\lambda,n} = E_{\lambda,n}\psi_{\lambda,n}$, where $H^{\bf B}$ is the Hamiltonian of the TQBC (type-I or II) in the presence of a magnetic field expressed in terms of the ladder operators, $E_{\lambda,n}$ is the eigenvalue of the LLs and $\psi_{\lambda,n}$ is the eigenvector. The band index $\lambda=-1,0,1$ indicates different groups of LLs; in particular, $\lambda=0$ corresponds to the LLs originating from the flat band. For type-I TQBCs, we obtain the LLs
\begin{equation}
\begin{cases}
E^I_{-1,n}=\tfrac{\hbar\omega_c}{2} \left(\tfrac{1}{2} -n- \sqrt{n(n-1)+1}\right), & \mbox{n=0, 1, 2, ...} \\
E^I_{0,n}=\tfrac{\hbar\omega_c}{2} \left(\tfrac{1}{2} -n+ \sqrt{n(n-1)+1}\right), & \mbox{n=2, 3, 4, ...} \\
E^I_{1,n}=\hbar\omega_c\left(n + \tfrac{1}{2}\right), & \mbox{n=0, 1, 2, ...} \\
\end{cases},\label{LLA}
\end{equation}
and for type-II TQBCs with $m_z=0$,
\begin{equation}
\begin{cases}
E^{II}_{-1,n}=-\hbar\omega_c\sqrt{n^2+n+1},  & \mbox{n=0, 1, 2, ...} \\
E^{II}_{0,n}=0,  & \mbox{n=-2, -1, 2, 3, ...}  \\
E^{II}_{1,n}=\hbar\omega_c\sqrt{n^2+n+1}, &\mbox{n=0, 1, 2, ...}
\end{cases},\label{LLB}
\end{equation}
where $\omega_c=eB/m$. As can be seen from Eqs.~(\ref{LLA}) and~(\ref{LLB}), for type-I TQBCs, the degeneracy of the flat band is lifted, displaying a series of anomalous nonzero-energy LLs; by contrast, for type-II TQBCs, the singular flat band in the gapless case has no anomalous LL structure. This is guaranteed by the peculiar structure of the eigenvector of the flat band Eq.~(\ref{VFB}) (with $m_z=0$), in which one component is always zero~\cite{Bohm-Jung2020Nature}. Note that the $n=0$ and $n=1$ levels are inexistent in the $\lambda=0$ groups for both the two systems~\cite{Malcolm2014PRB}.

For $H_{II}$ with $m_z\ne0$, the wave function can be written as $\psi^{II}_{\lambda, n}=(l_1|n-2\rangle, l_2|n\rangle, l_3|n+2\rangle)^T$. For $n=\{-1,-2\}$, we obtain $l_1=l_2=0$ and $l_3=1$ with $E^{II}_{n=-2}=E^{II}_{n=-1}=\hbar^2m_z/(2\sqrt{2}m)=\delta \hbar\omega_c$, where $\delta=\hbar m_z/(2\sqrt{2}eB)$ is dimensionless. For $n=\{0,1 \}$, one have $l_1=0$, with the energies being $E^{II}_{n=0,1}=\tfrac{1}{2}\hbar\omega_c \left(\delta \pm \sqrt{2(n+1)(n+2)+\delta^2}\right)$. The LLs for $n\ge 2$ are solved from the Hamiltonian
\begin{eqnarray}
H(n)=-\hbar \omega_c
\begin{pmatrix}
\delta & \frac{1}{\sqrt{2}}M_n & 0 \\
\frac{1}{\sqrt{2}}M_n & 0 & \frac{1}{\sqrt{2}}M_{n+2} \\
0 & \frac{1}{\sqrt{2}}M_{n+2} & -\delta
\end{pmatrix},
\label{Eq.LLN}
\end{eqnarray}
where $M_n=\sqrt{n(n-1)}$. Significantly, a constant $m_z$ gives the levels $E^{II}_{-1}$, $E^{II}_{-2}$ independent of the magnetic field ${\bf B}$, while, in order to get a scalable Hall conductivity (see Fig.~\ref{QHb}) we consider the $m_z \propto B$, so in Fig.~\ref{LLsb} (blue lines) the LLs have linear relation with the magnetic field ${\bf B}$. The blue levels also elucidate that in the gapped state of the type-II TQBC, the isolated flat band (IFB) has an anomalous Landau level structure. We highlight the $n=2$ levels in $\lambda=0$ groups by red lines for both types of TQBCs, as shown in Figs.~\ref{LLsa} and~\ref{LLsb}, to indicate that LLs of these flat bands get closer and closer to zero energy with increasing $n$.

An IFB is non-singular and its response to the magnetic field can be explained by considering a semiclassical $B$-linear quantum correction $\mu_{\lambda}({\bf k})B$ in the modified band structure~\cite{MingChe1996PRB}, where $\mu_{\lambda}({\bf k})$ is the orbital magnetic moment of the $\lambda$-th magnetic band in the $z$-direction. For a zero-energy IFB, one have
\begin{equation}
\mu_{\lambda}({\bf k}) = - \frac{e}{\hbar} {\rm Im} \langle \partial_{k_x} u_{\lambda}({\bf k})|H({\bf k})|\partial_{k_y} u_{\lambda}({\bf k})\rangle,
\end{equation}
with $|u_{\lambda}({\bf k})\rangle$ being the eigenvector of the IFB. In fact, the semiclassical correction corresponds to the interband couplings between the IFB and other bands~\cite{hwang2021NC}. The modified band dispersion of the zero-energy IFB is estimated as
\begin{equation}
E_{\lambda,{\bf B}}({\bf k}) = \mu_{\lambda}({\bf k})B = - \frac{eB}{\hbar} {\rm Im} \sum_{\lambda^\prime\neq {\lambda}}\varepsilon_{\lambda^\prime}({\bf k})\chi_{k_xk_y}^{\lambda^\prime \lambda}({\bf k}),
\end{equation}
in which
\begin{equation}
\chi_{k_xk_y}^{\lambda^\prime \lambda}({\bf k}) = \langle \partial_{k_x} u_{\lambda}({\bf k}) | u_{\lambda^\prime}({\bf k})\rangle \langle u_{\lambda^\prime}({\bf k}) | \partial_{k_y} u_{\lambda}({\bf k})\rangle
\end{equation}
is the fidelity tensor, $\varepsilon_{\lambda^\prime}({\bf k})$ is the energy of the $\lambda^\prime$-th band at zero magnetic field, and $\langle u_{\lambda^\prime}({\bf k})|\partial_i u_{\lambda}({\bf k})\rangle=A_i^{\lambda^\prime \lambda}({\bf k})$ indicates the cross-gap Berry connection between the $\lambda^\prime$-th and $\lambda$-th bands ($\lambda^\prime \neq \lambda$).

For the IFB in gapped type-II TQBCs, the calculation gives the modified band structure
\begin{equation}
E_{0,{\bf B}}({\bf k}) = - \hbar \omega_c \frac{2\sqrt{2}k^2m_z}{2k^4+m_z^2}.
\end{equation}
We note that $E^{max}_{0,{\bf B}}({\bf k})=E_{0,{\bf B}}(k=0)=0$ and $E^{min}_{0,{\bf B}}({\bf k})=E_{0,{\bf B}}\left(k=m_z/\sqrt[4]{2}\right)=-\hbar \omega_c$. These minimum and maximum values of $E_{0,{\bf B}}({\bf k})$ correspond to the lower and upper bounds for LLs of the IFB, respectively. However, this result is valid only when the band gap $E_{gap}$ between the IFB and its neighboring band at zero magnetic field is large enough, i.e., $E_{gap}\gg {\rm max}|E_{0,{\bf B}}({\bf k})|$.

\subsection{Unconventional quantum Hall effect}
Next, we calculate the Hall conductivities for the two types of TQBCs. The Hall conductivity at zero temperature is found via the Kubo formula~\cite{Tse.Wang-Kong2011PRB,Malcolm2014PRB},
\begin{equation}
\sigma_{xy}=\frac{ig}{2\pi\hbar l_B^2}\sum_{LL^\prime s}\frac{f_{\lambda n}-f_{\lambda^\prime n^\prime}}{(E_{\lambda n}-E_{\lambda^\prime n^\prime})^2} \langle \psi| J_x| {\psi}^\prime\rangle \langle {\psi}^\prime| J_y| \psi\rangle,
\label{HallC}
\end{equation}
where $g$ is the degeneracy factor, $J_{x(y)}=e\partial H/\partial P_{x(y)}$, $|\psi\rangle\equiv|\lambda n\rangle$ is the current operator, $|\psi^\prime\rangle \equiv |\lambda^\prime n^\prime\rangle$ are the LL eigenstates, $f_{\lambda n}$ is the Feimi distribution function and the index $LL^\prime s$ implies that the summation takes place over all initial and final Landau states. We neglect the spin degree of freedom and the valley degeneracy, so that $g=1$ in Eq.~(\ref{HallC}).
\begin{figure}[t]
  \centering
  \subfigure{\includegraphics[width=1.6in]{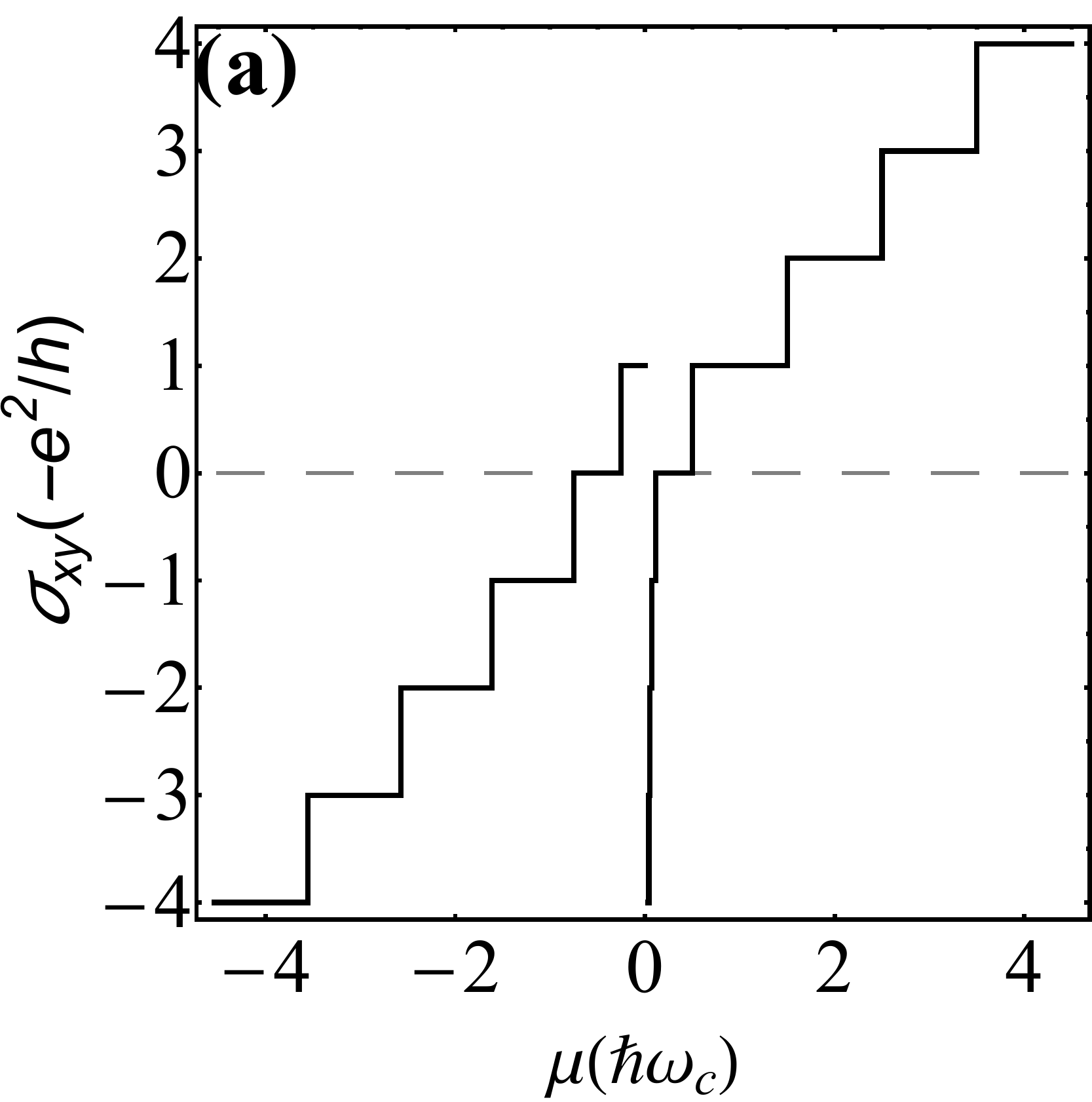}\label{QHa}}~~
  \subfigure{\includegraphics[width=1.6in]{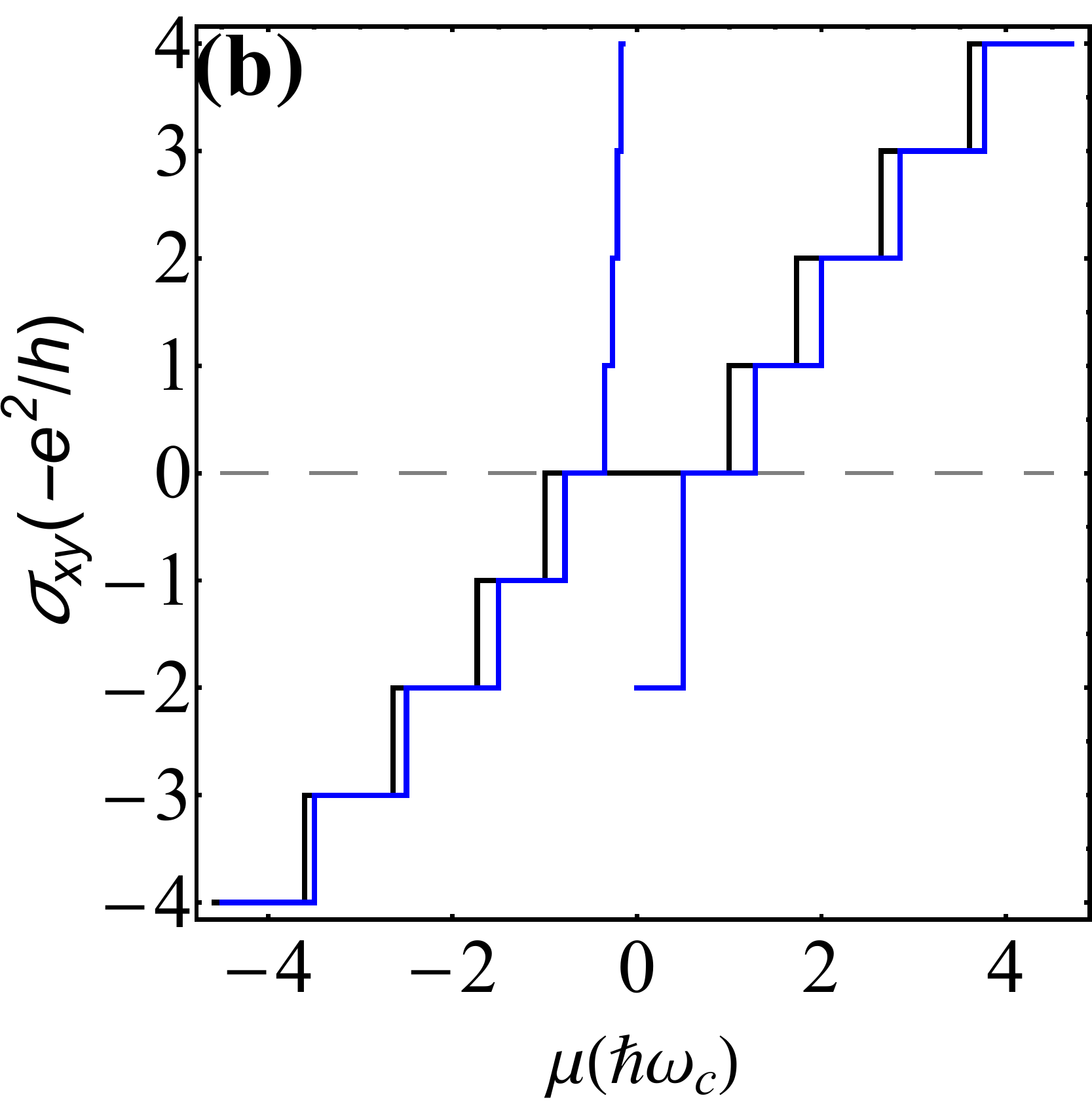}\label{QHb}}
  \caption{The Hall conductivities measured in units of $-e^2/h$ as a function of the chemical potential $\mu$ for type-I (a) and type-II (b) TQBCs. In (b), the black and blue lines correspond to $\delta=0$ and $\delta=0.5$, respectively.} \label{QH}
\end{figure}

For type-I TQBCs, the flat band generates a series of LLs near $E=0$. Therefore, when tuning chemical potential $\mu$ from positive toward 0, the Hall conductivity first decreases to zero and then exhibits an infinite ladder of plateaus. As the chemical potential is further moved across 0, the conductivity suddenly changes to $e^2/h$ and then decreases, as shown in Fig.~\ref{QHa}. The similar phenomenon can also be observed in type-II TQBCs when $m_z\neq0$, whose Hall conductivity is calculated numerically due to the fact that Eq.~(\ref{Eq.LLN}) can not be solved analytically. As shown by the blue lines in Fig.~\ref{QHb}, when increasing chemical potential $\mu$ from negative toward 0, an infinite ladder emerges in the Hall conductivity. Tuning $\mu$ further into a positive, the conductivity suddenly changes to $-2e^2/h$ owing to the degeneracy of $n=-1$ and $n=-2$ levels. For type-II TQBCs with $m_z=0$, the zero-energy LL is non-topological and has no contribution to the Hall conductivity. Therefore, the Hall conductivity is located at the zero-plateau when the chemical potential is near 0, as shown in Fig.~\ref{QHb} by the black lines. Actually, type-II TQBCs shows a similar behavior of Hall conductivity as pseudospin-1 DW fermions in both the gapless and gapped cases~\cite{Xu.Yong2017PRB}.

\section{Barrier tunneling of quasiparticles near TQBCs}\label{sec:Klein}
\begin{figure}[t]
  \centering
  \subfigure{\includegraphics[width=3.2in]{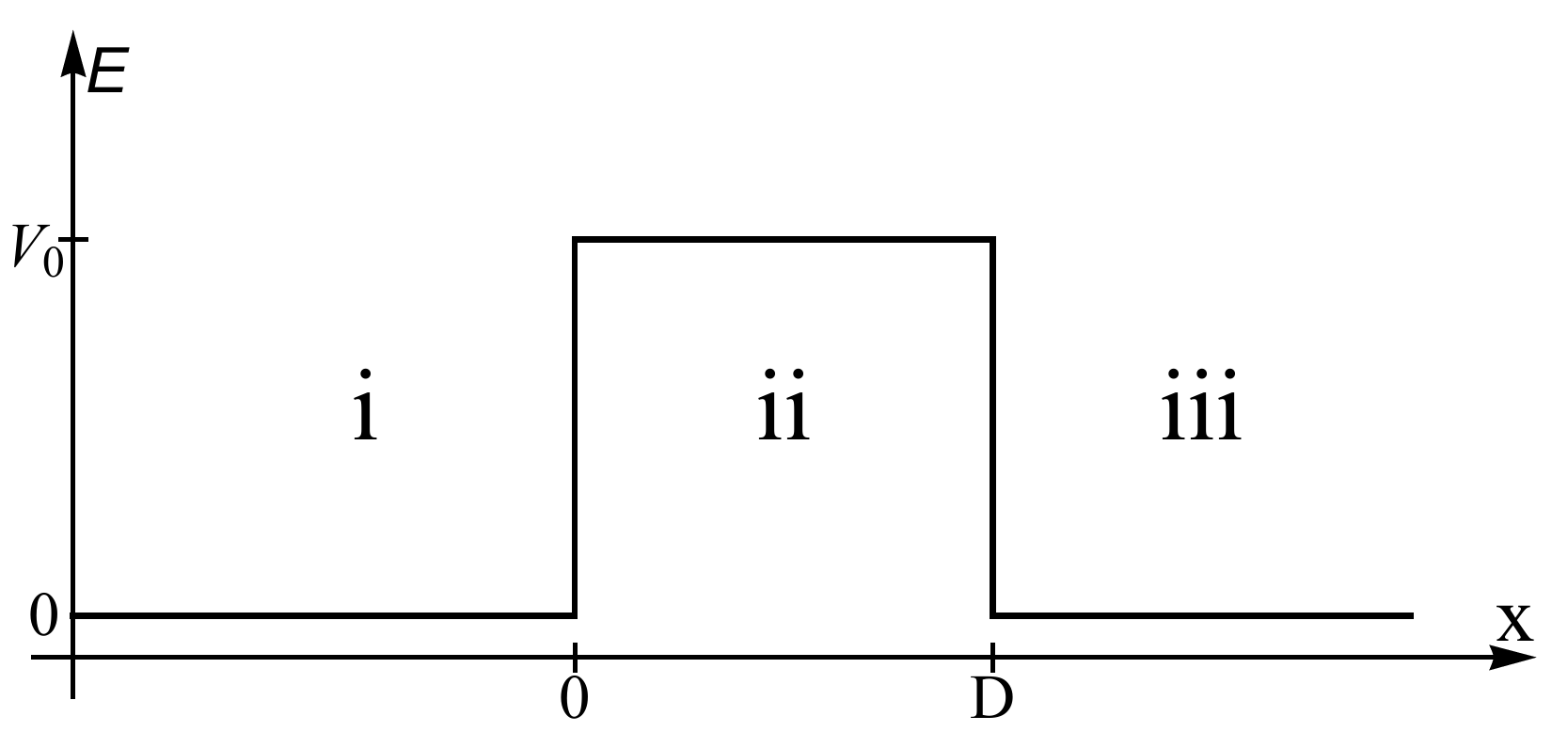}}
  \caption{Sketch of the scattering region with an electrostatic barrier of width $D$ and height $V_0$. The barrier is infinite in the $y$-direction.} \label{BaV}
\end{figure}

In contrast to the tunneling of Dirac fermions in graphene where Klein tunneling occurs only at normal incidence, the barrier transmission of pseudospin-1 DW fermions exhibits an all-angle perfect tunneling when the energy of incident electrons is equal to half the barrier height~\cite{Urban2011PRB}. Here, we address the problem of barrier tunneling of quasiparticles in the vicinity of TQBCs. The scattering region is shown in Fig.~\ref{BaV}, where an electrostatic barrier with height $V_0$ is at interval $[0, D]$ in the $x$-direction and extends infinitely in the $y$-direction.

\subsection{Barrier tunneling of quasiparticles near type-I TQBCs}
As discussed above, a type-I TQBC consists of two decoupled subsystems, a symmetry-protected QBC and a 2D free-electron band. For the sake of simplicity, here we assume the energy of incident particles in both the two subsystems to be $E=\hbar^2 k^2/2m$, with incident wavevector ${\bf k}$($k_x$, $k_y$). The wave function of model $H_I$ is written in the form $\psi^I(x,y)=\left[\varphi_L(x), \varphi_M(x), \varphi_R(x)\right]^Te^{ik_yy}$. First, we consider the case of $E<V_0$. In region (i) (see Fig.~\ref{BaV}), the wave function is composed of incoming and outgoing plane waves,
\begin{eqnarray}
\psi^I_1(x,y)=
\begin{pmatrix}
a e^{ik_xx} + b e^{-ik_xx} \\
a e^{ik_xx+2i\theta} + b e^{-ik_xx-2i\theta} \\
a^\prime e^{ik_xx}+b^\prime e^{-ik_xx}
\end{pmatrix}
e^{ik_yy}.
\label{WaveI1}
\end{eqnarray}
Inside the barrier where $0\leqslant x \leqslant D$, the wave function includes only evanescent waves,
\begin{eqnarray}
\psi^I_2(x,y)=
\begin{pmatrix}
c e^{\kappa_x x} + d e^{-\kappa_x x} \\
c \xi e^{\kappa_x x} + \tfrac{d}{\xi} e^{-\kappa_x x} \\
c^\prime e^{\kappa_x x}+d^\prime e^{-\kappa_x x}
\end{pmatrix}
e^{ik_yy}.
\label{WaveI2}
\end{eqnarray}
In region (iii) where $x>D$, the reflected waves vanish, so the wave function is written as
\begin{eqnarray}
\psi^I_3(x,y)=
\begin{pmatrix}
t e^{ik_xx}  \\
t e^{ik_xx+2i\theta} \\
t^\prime e^{ik_xx}
\end{pmatrix}
e^{ik_yy}.
\label{WaveI3}
\end{eqnarray}
In Eqs.~(\ref{WaveI1})$-$(\ref{WaveI3}), $\theta=\arctan{(k_y/k_x)}$ is the incident angle, $\hbar k_x=\sqrt{2mE}\cos{\theta}$, $\kappa_x=\sqrt{2m(V_0-E)/\hbar^2+k_y^2}$, $\xi=(\kappa_x-k_y)^2/(\kappa_x^2-k_y^2)$, and $t$ and $t^\prime$ are the transmission coefficients for the QBC subsystem and 2D free electrons, respectively. Integrating the eigenvalue equation $H_I\psi^I=E\psi^I$ over the small interval $x\in[-\epsilon,\epsilon]$ along the $x$-direction and letting $\epsilon$ eventually go to zero yields
\begin{subequations}
 \begin{align}
 \left[\varphi_L(x)+\varphi_M(x)\right]\big|^{\epsilon}_{-\epsilon} &=0, \label{BCAa}\\
 \partial_x \left[\varphi_L(x)+\varphi_M(x)\right]\big|^{\epsilon}_{-\epsilon}  &= 2k_y\varphi_L(x)\big|^{\epsilon}_{-\epsilon}, \label{BCAb} \\
 \varphi_R(x)\big|^{\epsilon}_{-\epsilon} &= 0,  \label{BCAc} \\
 \partial_x\varphi_R(x)\big|^{\epsilon}_{-\epsilon} &= 0  \label{BCAd}
 \end{align}\label{BCA}%
\end{subequations}
where $f(x)|^{\epsilon}_{-\epsilon}=f(\epsilon)-f(-\epsilon)$. At the barrier boundaries $x=0$ and $x=D$, both components of the wave function and their derivatives should satisfy the continuity conditions. As we can see in the wave function of model $H_I$, both the spinor components $\varphi_L$ and $\varphi_M$ are irrelevant to $\varphi_R$. As a consequence, particles in the free-electron band cannot go into the QBC subsystem via the Klein tunneling, and vice versa.
\begin{figure}[t]
  \centering
  \subfigure{\includegraphics[width=1.6in]{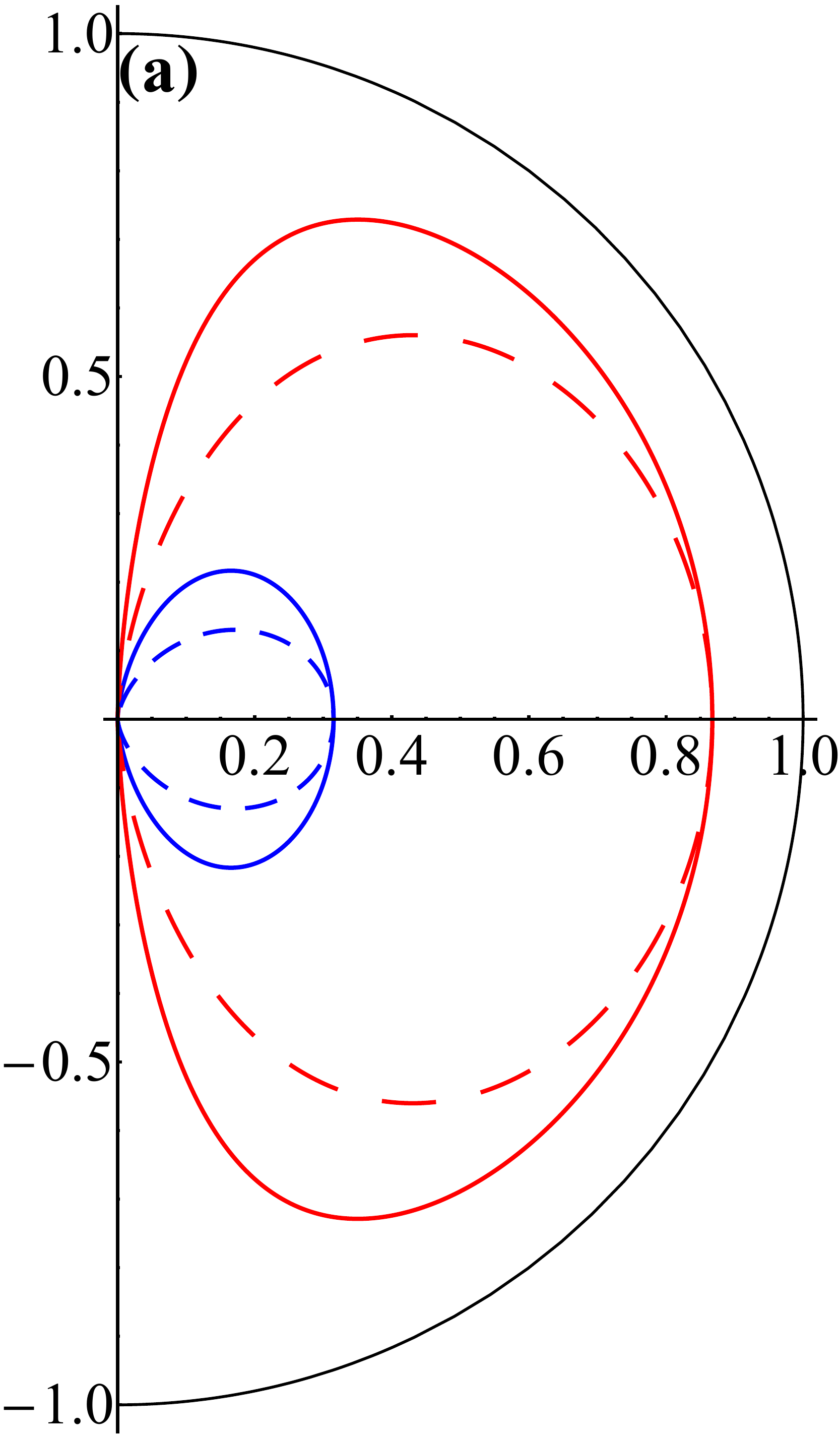}\label{KTc}}~~
  \subfigure{\includegraphics[width=1.6in]{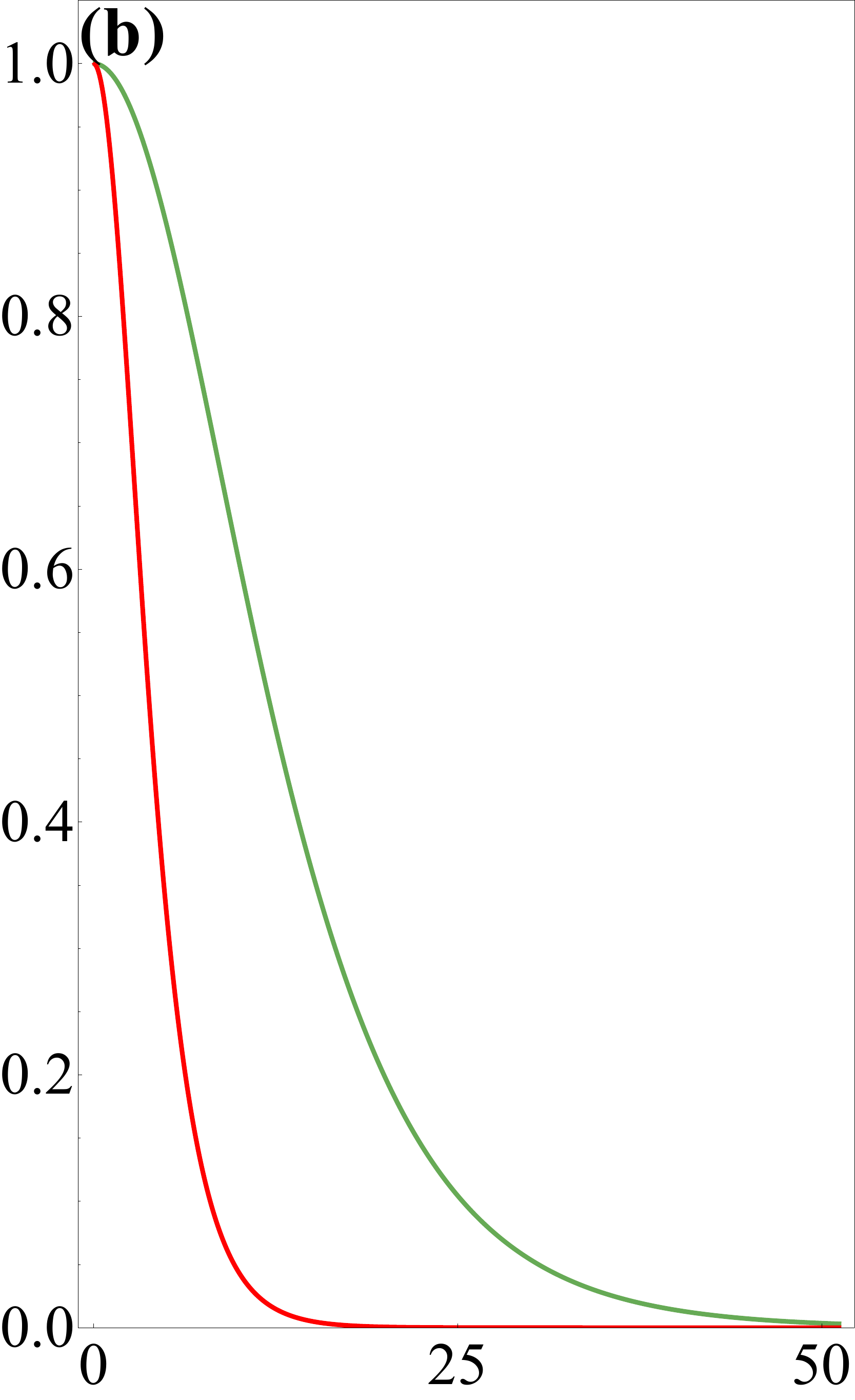}\label{KTDs}}
  \caption{Transmission probabilities $T$ through a 50-nm-wide barrier, with energy $E=15\mbox{meV}$ and barrier height $V_0=50\mbox{meV}$. (a) $T$ as a function of the incident angle for the QBC subsystem (solid) and 2D free electrons (dashed). The red lines are for $m=m_0/1000$ and the blue lines are for $m=m_0/100$. (b) $T$ as a function of the width $D$ of the barrier for normally incident particles. In this case, type-I and type-II TQBCs have equal $T$ under the same parameters. The red line is for $m=m_0$ and the green line is for $m=m_0/10$.} \label{KTA}
\end{figure}

Substituting the components of the wave function given in Eqs.~(\ref{WaveI1})$-$(\ref{WaveI3}) and their derivatives into Eq.~(\ref{BCA}), we can get eight linear equations, while only the four of them, corresponding to Eqs.~(\ref{BCAa}) and~(\ref{BCAb}), determine the tunneling of QBCs, and the other four equations, corresponding to Eqs.~(\ref{BCAc}) and~(\ref{BCAd}), give the transmission coefficient of the 2D free electrons. Finally, the transmission coefficient for the QBC subsystem is
\begin{equation}
t=\left[\frac{\Delta_{+}}{2} + \frac{\left[\kappa_x^2k^2 - (\kappa_x^2-k_y^2)^2\right]\cos\theta + \kappa_x^2k_y^2\sec\theta}{4i\kappa_x k(\kappa_x^2-k_y^2)}\Delta_{-} \right]^{-1},
\label{TrQB}
\end{equation}
where $\Delta_{+}=e^{\kappa_x D}+e^{-\kappa_x D}$ and $\Delta_{-}=e^{\kappa_x D}-e^{-\kappa_x D}$. For 2D free electrons, the transmission coefficient $t^\prime$ reads
\begin{equation}
t^\prime=\frac{4ik_x\kappa_x}{(\kappa_x+ik_x)^2e^{-\kappa_xD} - (\kappa_x-ik_x)^2e^{\kappa_xD}}.
\label{TrND}
\end{equation}
The transmission probabilities $T=|t|^2$ as a function of the incident angle for the QBC subsystem and 2D free electrons are shown in Fig.~\ref{KTc}. The two subsystems show similar tunneling properties, that is, the transmission probabilities decay exponentially with the barrier width for any incident angle, and are inversely proportional to the effective mass of quasiparticles. We plot the transmission probability for $m=m_0/1000$ and $m=m_0/100$ in Fig.~\ref{KTc}. In particular, for normally incident particles, i.e., $k_y=0$, the transmission coefficients in Eqs.~(\ref{TrQB}) and~(\ref{TrND}) coincide.

In the QBC subsystem, the flat band with $E_k=0$ prompts us to study a special case corresponding to $E=V_0$. Firstly, we consider the normally incident particles, the wave function in the barrier is
\begin{eqnarray}
\psi^I_2(x) &= \sum\limits_{q}\alpha_q
\begin{pmatrix}
e^{iqx}  \\
-e^{iqx}
\end{pmatrix}
+ \sum\limits_{\varkappa}\beta_\varkappa
\begin{pmatrix}
e^{\varkappa x}  \\
-e^{\varkappa x}
\end{pmatrix}
+
\begin{pmatrix}
Lx+C  \\
L^\prime x+C^\prime
\end{pmatrix},
\label{FWI2}
\end{eqnarray}
Applying the continuity conditions~(\ref{BCA}), that gives the transmission probability
\begin{eqnarray}
T=\frac{4}{4+k_x^2 D^2}.
\label{TEV}
\end{eqnarray}
At $E=V_0$ and finite momentum $k_y$, the wave function~(\ref{FWI2}) should be replaced with
\begin{eqnarray}
\psi^I_2(x) &= \sum\limits_{q}\alpha_q
\begin{pmatrix}
e^{iqx}  \\
-e^{iqx+2i\theta_q}
\end{pmatrix}
+ \sum\limits_{\varkappa}\beta_\varkappa
\begin{pmatrix}
e^{\varkappa x}  \\
-\xi_{\varkappa}e^{\varkappa x}
\end{pmatrix}
+
\begin{pmatrix}
Le^{k_yx}  \\
L^\prime e^{-k_yx}
\end{pmatrix},\label{Fwm}
\end{eqnarray}
where $\theta_q=\arctan{(k_y/q)}$ and $\xi_{\varkappa}=(\varkappa-k_y)^2/(\varkappa^2-k_y^2)$. Through a tedious but straightforward calculation, we eventually get the transmission probability $T=0$.

\begin{figure}[t]
  \centering
  \subfigure{\includegraphics[width=1.6in]{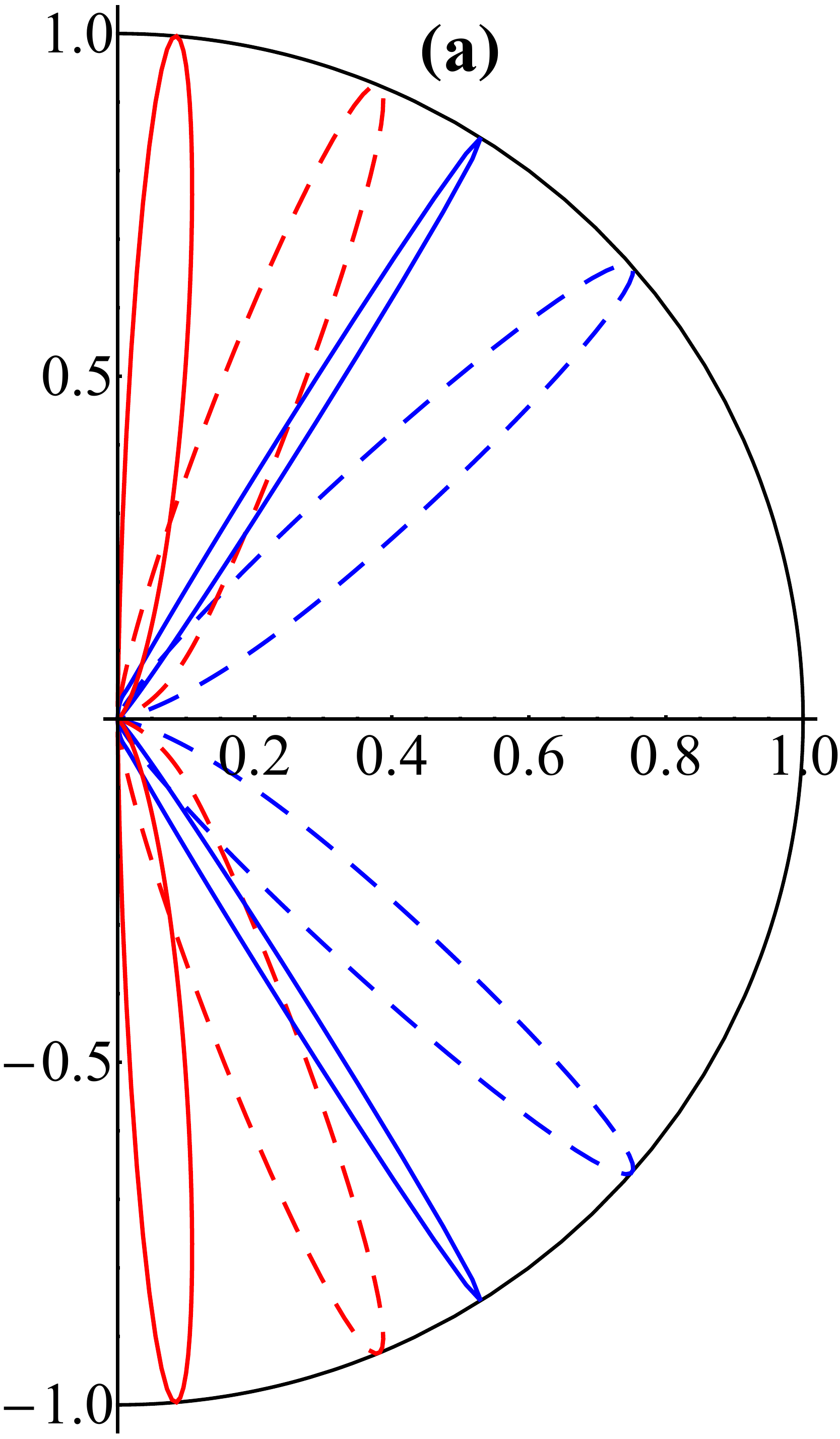}\label{KTt}}~~
  \subfigure{\includegraphics[width=1.6in]{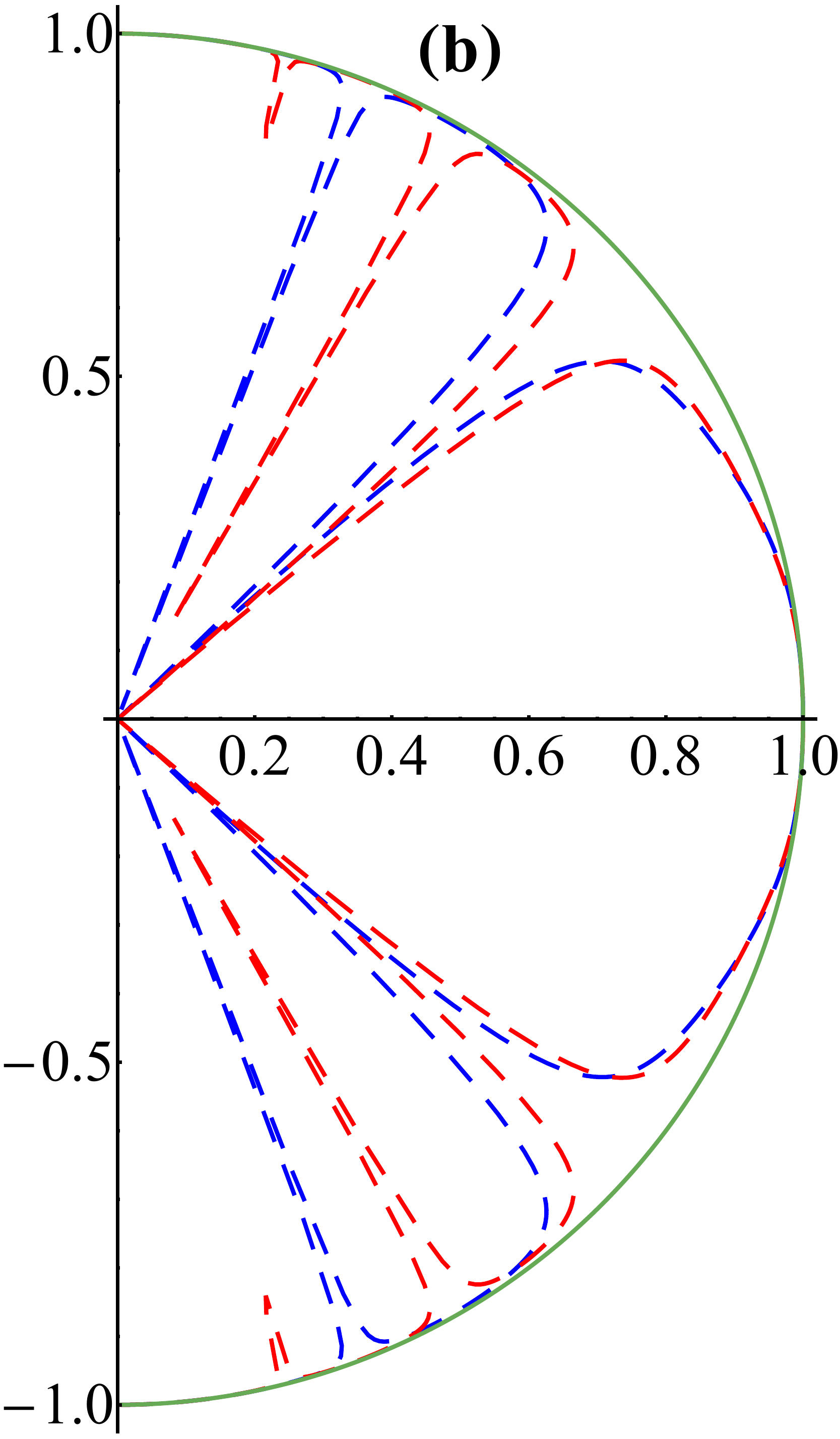}\label{KTr}}
  \caption{Transmission probabilities $T$ (a) and reflection probabilities $R$ (b) through a 50-nm-wide barrier as a function of the incident angle for bilayer graphene (dashed) and type-II TQBC (solid), the blue lines are for $V_0=50$meV and the red lines are for $V_0=100$meV. The energies are $E=15$meV (a) and $E=V_0/2$ (b), respectively. In (b), the reflection probabilities of type-II TQBC under different parameters are equal and represented by the green line.} \label{KTB}
\end{figure}

\subsection{Barrier tunneling of quasiparticles near type-II TQBCs}
Now we investigate the tunneling of quasiparticles near type-II TQBCs. In the constant potential $V_i$, the wave function, in the form of $\psi^{II}(x,y)=\left[\varphi_A(x),\varphi_H(x),\varphi_B(x)\right]^Te^{ik_yy}$, includes not only propagating waves but also evanescent waves,
\begin{eqnarray}
\begin{split}
\psi^{II}_i(x) &= a_ie^{ik_{ix}x}
\begin{pmatrix}
e^{-2i\theta_i} \\
\sqrt{2}s_i \\
e^{2i\theta_i}
\end{pmatrix}
+ b_ie^{-ik_{ix}x}
\begin{pmatrix}
e^{2i\theta_i} \\
\sqrt{2}s_i \\
e^{-2i\theta_i}
\end{pmatrix} \\
&-c_ie^{\kappa_{ix}x}
\begin{pmatrix}
\xi_i^{-1} \\
-\sqrt{2}s_i \\
\xi_i
\end{pmatrix}
-d_ie^{-\kappa_{ix}x}
\begin{pmatrix}
\xi_i \\
-\sqrt{2}s_i \\
\xi_i^{-1}
\end{pmatrix},
\end{split}
\end{eqnarray}
where $i$ takes on the values 1, 2 and 3 corresponding to the three scattering regions (i), (ii) and (iii), respectively; $s_i=\mbox{sgn}(V_i-E)$, $\theta_i=\arctan{(k_y/k_{ix})}$, $\hbar k_{ix}=\sqrt{2m|E-V_i|}\cos{\theta_i}$, $\kappa_{ix}=\sqrt{k_{ix}^2+2k_y^2}$ and $\xi_i=(\sqrt{1+\sin^2{\theta_i}}-\sin{\theta_i})^2$. Obviously, we should set $d_1=0$ for $x<0$ and $b_3=c_3=0$ for $x>0$ due to the finiteness of the wave function. The continuity conditions are obtained by integrating the eigenvalue equation $H_{II}\psi^{II}=E\psi^{II}$ in the same way as Eq.~(\ref{BCA}),
\begin{subequations}
 \begin{align}
 \left[\varphi_A(x)+\varphi_B(x)\right]\big|^{\epsilon}_{-\epsilon} &= 0, \label{BCBa} \\
 \partial_x\left[\varphi_A(x)+\varphi_B(x)\right]\big|^{\epsilon}_{-\epsilon}  &= 2k_y\left[\varphi_A(x)-\varphi_B(x)\right]\big|^{\epsilon}_{-\epsilon}, \label{BCBb} \\
 \varphi_H(x)\big|^{\epsilon}_{-\epsilon} &= 0, \label{BCBc} \\
 \partial_x\varphi_H(x)\big|^{\epsilon}_{-\epsilon} &= 0. \label{BCBd}
 \end{align}\label{BCB}%
\end{subequations}
Similarly, components of the wave function of model $H_{II}$ and their derivatives have to satisfy Eq.~(\ref{BCB}) by matching up coefficients $a_i$, $b_i$, $c_i$ and $d_i$.

In the continuity conditions~(\ref{BCB}), we also get eight linear equations, which determine the tunneling of quasiparticles near type-II TQBC. While the expression of the transmission coefficient for this case is too complicated, we present only the numerical results. Besides, since model $H_{II}$ is an extension of the effective model of QBCs in bilayer graphene, we also plot the transmission probability for bilayer graphene under the same parameters, as shown in Fig.~\ref{KTB}. For $E<V_0/2$, the tunneling of type-II TQBC is highly anisotropic with respect to the incident angle, see Fig.~\ref{KTt}, and displays transmission probability approaching unity at some angles, which is similar to the case of bilayer graphene.

However, we find that the tunneling of type-II TQBC shows a dramatic difference compared with the case of bilayer graphene when $E=V_0/2$ -- the former hosts an all-angle perfect reflection, while the latter shows again pronounced transmission resonances at some incident angles. For clarity, we plot the reflection probability $R=1-T$ as a function of the incident angle, as shown in Fig.~\ref{KTr}. The reflection probability for type-II TQBC, the green line, always approaches unity for all incident angles. With the same parameters, the reflection probability for bilayer graphene, the dashed lines, drops to zero at some angles. In fact, in order to investigate the special case of $E=V_0/2$, we performed a series of numerical calculations at different barrier heights and widths (the barrier heights are within the range of 20meV to 200meV, and the widths 20nm to 200nm). Take the case of $V_0=20$meV as an example, one finds that there is no significant transmission amplitude that can be observed experimentally for any incident angle when the barrier width $D>D_c \approx 25$nm, and for a higher barrier, the critical width $D_c$ decreases. In the calculations we take $m=m_0$.

We also calculate the transmission of normally incident particles for type-II TQBC, and get the same transmission coefficient as the QBC subsystem in type-I TQBC, as well as 2D free electrons and bilayer graphene~\cite{Katsnelson2006NatureP}, which can be obtained via Eq.~(\ref{TrND}) with $k_y=0$. Their transmission probability as a function of width $D$ of the barrier are shown in Fig.~\ref{KTDs}. Significantly, for type-I TQBC, the wave functions of both the QBC subsystem and 2D free electrons include only evanescent waves inside the barrier, but for type-II TQBC and bilayer graphene, there are plenty of electronic states inside the barrier.

Finally, we consider the case of $E=V_0$. For normally incident particles, we have the wave function in the barrier
\begin{eqnarray}
\psi^{II}_2(x) &=
\begin{pmatrix}
Lx+C \\
\eta x+\eta^\prime \\
L^\prime x+C^\prime
\end{pmatrix}
+ \sum\limits_{q}\alpha_q
\begin{pmatrix}
1 \\
0  \\
-1
\end{pmatrix}
e^{iqx} - \sum\limits_{\varkappa}\beta_\varkappa
\begin{pmatrix}
1 \\
0 \\
-1
\end{pmatrix}
e^{\varkappa x}.
\end{eqnarray}
In the continuity conditions~(\ref{BCB}), the wave function $\psi^{II}_2$ produces the same transmission probability as Eq.~(\ref{TEV}), which decays with the barrier width. That is different from the case of pseudospin-1 DW fermions, in which the flat band contributes the transmission probability of approaches unity when $E=V_0$ and $k_y=0$~\cite{Urban2011PRB}. At $E=V_0$ and $k_y\neq 0$, the wave function in the barrier is written as
\begin{eqnarray}
\psi^{II}_2(x) &=
\begin{pmatrix}
Le^{k_yx} \\
0 \\
L^\prime e^{-k_yx}
\end{pmatrix}
+ \sum\limits_{q}\alpha_q
\begin{pmatrix}
e^{iqx-2i\theta_q} \\
0  \\
-e^{iqx+2i\theta_q}
\end{pmatrix}
 - \sum\limits_{\varkappa}\beta_\varkappa
\begin{pmatrix}
\xi_{\varkappa}^{-1}e^{\varkappa x} \\
0 \\
-\xi_{\varkappa}e^{\varkappa x}
\end{pmatrix},
\end{eqnarray}
where $\theta_q$ and $\xi_{\varkappa}$ have been defined in Eq.~(\ref{Fwm}). Because of the vanishing $\varphi_H$ component, we directly get the transmission probability $T=0$.

\section{Summary}\label{sec:summary}
We have studied the transport properties of two types of TQBCs. The LLs and Hall conductivity in magnetic fields are calculated, and the tunneling of the nodal quasiparticles through electrostatic barriers is investigated. The first system, type-I TQBC, is composed of a symmetry-protected QBC and a free-electron band, and can be realized in the AA-stacked bilayer square-octagon lattice. In a magnetic field, the singular flat band displays an anomalous LL structure that produces an infinite ladder of Hall plateaus in the Hall conductivity when the chemical potential is tuned toward zero. Under perturbations, the QBC subsystem in type-I TQBC can split into two Dirac points when the $C_4$ symmetry of the square-octagon lattice is broken down to $C_2$, and the free-electron band touches the middle band on an accidental nodal loop. Compared with the Klein tunneling in bilayer graphene, the transmission probability of QBCs here decays exponentially with the barrier width for any incident angle, owing to the lack of propagating waves inside the barrier.

The other model, type-II TQBC, is a pseudospin-1 generalization of the effective model of the QBC in bilayer graphene. In the presence of a magnetic field, in the gapped case, the IFB also exhibits an anomalous LL structure which induces an infinite ladder in its Hall conductivity when tuning the chemical potential $\mu$ toward zero. But in the gapless case, the zero-energy LL of type-II TQBC is non-topological and has no contribution to the Hall conductivity, so that a zero-plateau is present when $\mu$ is near zero. Under perturbations, the second type of TQBCs may split into several linear pseudospin-1 DW fermions in a similar way as the splitting of QBCs~\cite{Mikitik2008PRB}, and the total winding number is conserved. Through an electrostatic barrier with $V_0>2E$, the Klein tunneling in type-II TQBC shows pronounced transmission resonances at some incident angles, while when the energy of incident particles approaches half of the barrier height, the tunneling hosts an all-angle perfect reflection for a sufficiently wide barrier.

\acknowledgments
This work was supported by Guangdong Basic and Applied Basic Research Foundation (Grant No. 2021B1515130007), Shenzhen Natural Science Fund (the Stable Support Plan Program 20220810130956001), and National Natural Science Foundation of China (Grant Nos. 12004442, 92165204, and 11974432).

\bibliography{quadratic}

\bibliographystyle{apsrev4-2}

\end{document}